\documentclass{revtex4}
\usepackage{amsthm,amssymb,amsmath}				
\usepackage{graphicx,comment}
\usepackage{float}   
 \usepackage{xcolor}

\begin{document}

\title{PDM damped-driven oscillators: exact solvability, classical states crossings, and self-crossings}
\author{Omar Mustafa}
\email{omar.mustafa@emu.edu.tr}
\affiliation{Department of Physics, Eastern Mediterranean University, G. Magusa, north
Cyprus, Mersin 10 - Turkey,\\
Tel.: +90 392 6301378; fax: +90 3692 365 1604.}

\begin{abstract}
\textbf{Abstract:}\ Within the standard Lagrangian and Hamiltonian setting, we consider a position-dependent mass (PDM) classical particle performing a damped driven oscillatory (DDO) motion under the influence of a conservative harmonic oscillator force field $V\left( x\right) =\frac{1}{2}\omega ^{2}Q\left( x\right) x^{2}$ and subjected to a Rayleigh dissipative force field $\mathcal{R}\left( x,\dot{x}\right) =\frac{1}{2}b\,m\left( x\right) \dot{x}^{2}$ in the presence of an external periodic (non-autonomous) force $F\left( t\right) =F_{\circ }\,\cos \left( \Omega t\right) $. Where, the correlation between the coordinate deformation $\sqrt{Q(x)}$ and the velocity deformation $\sqrt{m(x)}$ is governed by a point canonical transformation $q\left( x\right) =\int \sqrt{m\left( x\right) }dx=\sqrt{%
Q\left( x\right) }x$. Two illustrative examples are used: a non-singular PDM-DDO, and a power-law PDM-DDO models. Classical-states $\{x(t),p(t)\}$ crossings are analysed and reported. Yet, we observed/reported that as a classical state $\{x_{i}(t),p_{i}(t)\}$ evolves in time it may cross itself at an earlier and/or a latter time/s.


\end{abstract}

\maketitle

\section{Introduction}

The exact solvability of the harmonic oscillator dynamical equation offers
not only an impressive pedagogical implementations but also a
reflection/mapping of its exact solution into the more realistic damped harmonic 
oscillators (e.g., \cite{Symon 1972,Chandrasekar 2006,Chandrasekar
2012,Chandrasekar-PRE 2005,Carinena Ranada Sant 2004,Mustafa 2015,Mustafa
2019,Mustafa 2021,Mustafa1 2021} and related references cited therein). For
example, the linear harmonic oscillator equation%
\begin{equation}
\ddot{U}+\omega ^{2}U=0\;\Longleftrightarrow \;U\left( t\right) =a\cos
\left( \omega t+\phi \right) +b\sin \left( \omega t+\phi \right) ,
\label{SHO}
\end{equation}%
would transform into a damped harmonic oscillator (DHO) equation%
\begin{equation}
\ddot{y}+2\eta \Omega \;\dot{y}+\Omega ^{2}y=0;\;\omega =\Omega \sqrt{1-\eta
^{2}},\eta \leq 1,  \label{DHO1}
\end{equation}%
through a nonlocal transformation of the form%
\begin{equation}
U\left( t\right) =e^{\eta \Omega t}y\left( t\right) \Longleftrightarrow
y\left( t\right) =e^{-\eta \Omega t}\left[ a\cos \left( \omega t+\phi
\right) +b\sin \left( \omega t+\phi \right) \right] .  \label{U-q mapping1}
\end{equation}%
This is just one example among so many discussed in more details by
Chandrasekar et al. \cite{Chandrasekar 2006} and some other nonlinear
oscillators (including position-dependent mass (PDM) ones) discussed in the
sample of references 
\cite{Chandrasekar 2012,Chandrasekar-PRE 2005,Carinena Ranada Sant 2004,Mustafa 2015,Mustafa 2019,
Mustafa 2021,Mustafa1 2021,M-L 1974,Tiwari 2013,Lak-Chand 2013,Pradeep 2009,Chand-Lak 2007}
. Oscillators (damped or undamped) find their applicability in many fields
of interest like the search for gravitational waves, laser cooling of atoms,
medical physics studies, etc (for more details on this issue, the reader may
refer to Marmolejo et al. \cite{Javier 2020}, and references cited therein).
Moreover, a classical particle subjected to a conservative harmonic
oscillator potential force field in the vicinity of damping and driving
forces (referred to as a damped driven oscillator (DDO) \cite{Symon 1972})
allows the resonance phenomenon to emerge (i.e., the frequency $\Omega $ of
the applied driving force matches the frequency $\omega $ of the simple harmonic oscillator,
as in equation (\ref{L(q)-damped-driven}) below). The resonance phenomena
are used in the study of classical mechanics, electromagnetism, optics,
acoustics, etc (e.g., \cite{Javier
2020,Ortiz1,Ortiz2,Ortiz3,Kharkongor,Saikia} and references cited therein). 

On the other hand, it has been asserted that if the coordinate $y$ in the
dynamical equation (\ref{DHO1}) is transformed/deformed in such a way that $%
y\longrightarrow \sqrt{Q\left( x\right) }x$, then the velocity would be
transformed/deformed in a completely different manner so that $\dot{y}%
\longrightarrow \sqrt{m\left( x\right) }\dot{x}$. This would keep the longstanding \emph{%
gain-loss balance correlation} between the kinetic and potential energies of the system
and, consequently, the total energy remains an integral of motion (i.e., the standard structure the textbook Lagrangians/Hamiltonians for
conservative systems of course). Such a transformation/deformation recipe would in effect introduce the so
called position-dependent effective mass concept  (or PDM in short) into
classical and quantum mechanics (c.f., e.g., sample of references 
\cite{Chandrasekar 2012,Chandrasekar-PRE 2005,Carinena Ranada Sant 2004,Mustafa 2015,
Mustafa 2019,Mustafa 2021,Mustafa1 2021,M-L 1974,Tiwari 2013,Lak-Chand 2013,Pradeep 2009,
Chand-Lak 2007,da Costa1 2020,Khlevniuk 2018,Mustafa Algadhi 2019,Quesne 2015,Mustafa Phys.Scr.
 2020,Ranada 2016,Carinena Herranz 2017,Bagchi Ghosh 2013,Mustafa 2020,dos Santos 2021,Mustafa arXiv,
Nabulsi1 2020,Nabulsi2 2020,Nabulsi3 2021,Nabulsi4 2021,Nabulsi5 2021,Quesne 2004,Quesne 2019,Mustafa arXiv1}
). It is, therefore, interesting to study and investigate PDM settings on such DDOs. To the best of our knowledge, the current PDM-DDO proposal has never been reported elsewhere.

In the current methodical proposal, we recycle/recollect (in section 2) the mathematical preliminaries for a constant mass, $m_{\circ}=1$, classical particle performing a damped driven oscillatory motion  under the influence of a conservative harmonic oscillator force field $V\left( q\right) =\frac{1}{2}\omega ^{2}q^{2}$, a Rayleigh dissipative force field $\mathcal{R}\left( \dot{q}\right) =\frac{1%
}{2}b\,\dot{q}^{2}$, and in the presence of an external periodic non-autonomous force $F\left( t\right) =F_{\circ }\,\cos \left( \Omega t\right)$. This would, in effect,  make our proposal self-contained. In section 3, we use a point canonical transformation, (\ref{PCT1}) below, and report the corresponding  PDM-Lagrangian as well as the corresponding PDM dynamical equation for the DDO. In the same section, we report our results for two PDM illustrative examples: a non-singular Mathews-Lakshmanan type \cite{M-L 1974}  PDM and a power-law one. Our concluding remarks are given in section 4.

\section{Damped driven oscillator: preliminaries recollected}

Consider a classical particle moving under the influence of a conservative oscillator force field $V\left( q\right) =\frac{1}{2}\omega ^{2}q^{2}$, a Rayleigh dissipative force field $\mathcal{R}\left( \dot{q}\right) =\frac{1%
}{2}b\,\dot{q}^{2}$, and in the presence of an external periodic non-autonomous force $F\left( t\right) =F_{\circ }\,\cos \left( \Omega t\right) $. Then the standard Lagrangian describing this particle is given by%
\begin{equation}
L\left( q,\dot{q},t\right) =\frac{1}{2}\dot{q}^{2}-\frac{1}{2}\omega
^{2}q^{2}.  \label{L(q)-damped-driven}
\end{equation}%
Under such dissipative and periodic forces settings, the Euler-Lagrange
equation of motion reads%
\begin{equation}
\frac{d}{dt}\left( \frac{\partial L}{\partial \dot{q}}\right) -\frac{%
\partial L}{\partial q}+\frac{\partial \mathcal{R}}{\partial \dot{q}}%
=F\left( t\right) \Longleftrightarrow \ddot{q}\left( t\right) +2\mathcal{%
\eta }\omega \,\dot{q}\left( t\right) +\omega ^{2}q\left( t\right) =F_{\circ
}\,\cos \left( \Omega t\right) ,  \label{q-DD oscillator eq}
\end{equation}%
where $\,\mathcal{\eta }=b/2\omega $ is the damping ratio. This dynamical equation is identified as the damped-driven oscillator (DDO) equation of motion. The solution of its homogeneous part (usually called the complementary or transient solution when $F_{\circ }=0$) is given by%
\begin{equation}
q\left( t\right) =e^{-\omega \mathcal{\eta }t}\left[ A_{t}\cosh \left( \beta
t\right) +B_{t}\sinh \left( \beta t\right) \right] \,;\,\beta =\omega \sqrt{%
\mathcal{\eta }^{2}-1}.  \label{q(t) general}
\end{equation}%
Where the values of $\eta $ identify the nature of damping so that one uses $%
\eta <1$ for under-damping, $\eta =1$ for critical-damping, and $\eta >1$ for over-damping. Moreover, under the assumption that $q\left( 0\right)
=q_{\circ }\neq 0$ and $\dot{q}\left( 0\right) =\dot{q}_{\circ }\neq 0$ the solution ( \ref{q(t) general}) reduces to%
\begin{equation}
q\left( t\right) =A_{t} e^{-\omega \mathcal{\eta }t}\cosh \left( \beta t\right) ,
\label{q(t) for F0=0}
\end{equation}%
where $B_{t}=0$ to avoid imaginary settings for $\dot{q}\left( 0\right)=\dot{q}_{\circ }\neq 0\in \mathbb{R}$ (i.e., $\beta =i\omega \sqrt{1-\mathcal{\eta }^{2}};$ for the under-damping case $\mathcal{\eta }<1$). As such, one may suggest that the general solution for (\ref{q-DD oscillator eq}) is of the form $q\left(
t\right) =Ae^{-\omega \mathcal{\eta }t}\cosh \left( \beta t\right) +\mathcal{F}_s\left( t\right) $, where $$\mathcal{F}_s\left( t\right)=C_s e^{i(\Omega t-\delta)} $$ is the so called steady state solution  \cite{Symon 1972}. When this assumption is plugged in (\ref{q-DD oscillator eq}), it yields%
\begin{equation}
\mathcal{\ddot{F}}_{s}\left( t\right) +2\mathcal{\eta }\omega \,\mathcal{\dot{F}}_{s}\left( t\right) +\omega ^{2}\mathcal{F}_{s}\left( t\right) =F_{\circ }\,\cos
\left( \Omega t\right) \Longleftrightarrow \mathcal{F}_{s}\left( t\right)
=C_s\,\cos \left( \Omega t-\delta \right) .  \label{steady state soluition}
\end{equation}
Then the general solution for the DDO dynamical equation \ref{q-DD oscillator eq} is given by 
\begin{equation}
q\left( t\right) =A_{t}\;e^{-\omega \mathcal{\eta }t}\cosh \left( \beta
t\right) +C_s\,\cos \left( \Omega t-\delta \right) ,
\label{q(t)-general solution}
\end{equation}%
where 
\begin{equation}
C_s=C_s\left( \Omega \right) =\frac{F_{\circ }}{\sqrt{\left[ \omega ^{2}-\Omega
^{2}\right] ^{2}+4\mathcal{\eta }^{2}\omega ^{2}\Omega ^{2}}}.
\label{DDO-amplitude}
\end{equation}%
Moreover, with $\cos(\delta)=C_s(\omega^2-\Omega^2)/F_{\circ}$ and $\sin(\delta)=2C_s\eta\omega\Omega/F_{\circ})$ one obtains%
\begin{equation}
\delta =\arctan \left( \frac{2\mathcal{\eta }\omega \,\Omega }{\omega
^{2}-\Omega ^{2}}\right) =\frac{\pi }{2}-\arctan \left( \frac{\omega
^{2}-\Omega ^{2}}{2\mathcal{\eta }\omega \,\Omega }\right) .
\label{phase shift}
\end{equation}%
The phase shift $\delta $ measures the phase lag between the external force $F(t)$ and the system's response (with a time lag $\tau =\delta/\Omega$). Moreover, one may find the amplitude resonance frequency by requiring that $dC\left( \Omega \right) /d\Omega =0$ to yield $
\Omega _{R}=\omega \sqrt{1-2\eta ^{2}}$.

\section{Damped driven Oscillator: PDM-counterpart}

A point canonical transformation of the form%
\begin{equation}
q\longrightarrow q\left( x\right) =\int \sqrt{m\left( x\right) }dx=\sqrt{%
Q\left( x\right) }x\Longleftrightarrow \sqrt{m\left( x\right) }=\sqrt{%
Q\left( x\right) }\left( 1+\frac{Q^{\prime }\left( x\right) }{2Q\left(
x\right) }x\right) ,  \label{PCT1}
\end{equation}%
would yield%
\begin{equation}
\dot{q}=\sqrt{Q\left( x\right) }\left( 1+\frac{Q^{\prime }\left( x\right) }{%
2Q\left( x\right) }x\right) \dot{x}\Longleftrightarrow \dot{q}=\sqrt{m\left(
x\right) }\dot{x}.  \label{q-dot}
\end{equation}%
Under such settings, the DDO-Lagrangian (\ref{L(q)-damped-driven}) would
transform into a PDM-DDO one so that%
\begin{equation}
L\left( q,\dot{q},t\right) =\frac{1}{2}\dot{q}^{2}-\frac{1}{2}%
\omega ^{2}q^{2}\Longleftrightarrow L\left( x,\dot{x},t\right) =%
\frac{1}{2}m\left( x\right) \dot{x}^{2}-\frac{1}{2}\omega ^{2}Q\left( x\right) x^{2}.  \label{PDM-DDO-L}
\end{equation}%
\begin{figure}[!ht]  
\centering
\includegraphics[width=0.3\textwidth]{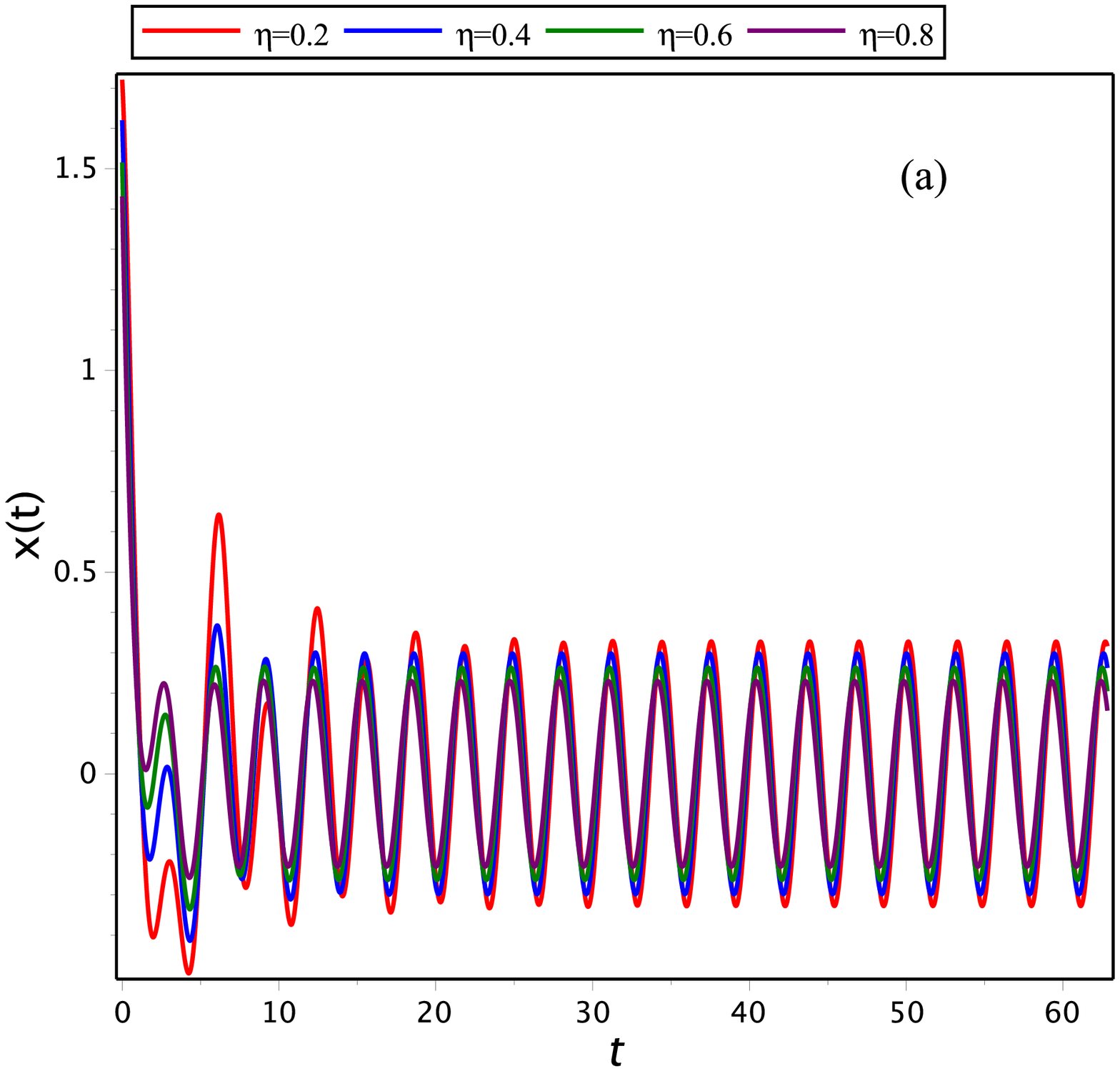}
\includegraphics[width=0.3\textwidth]{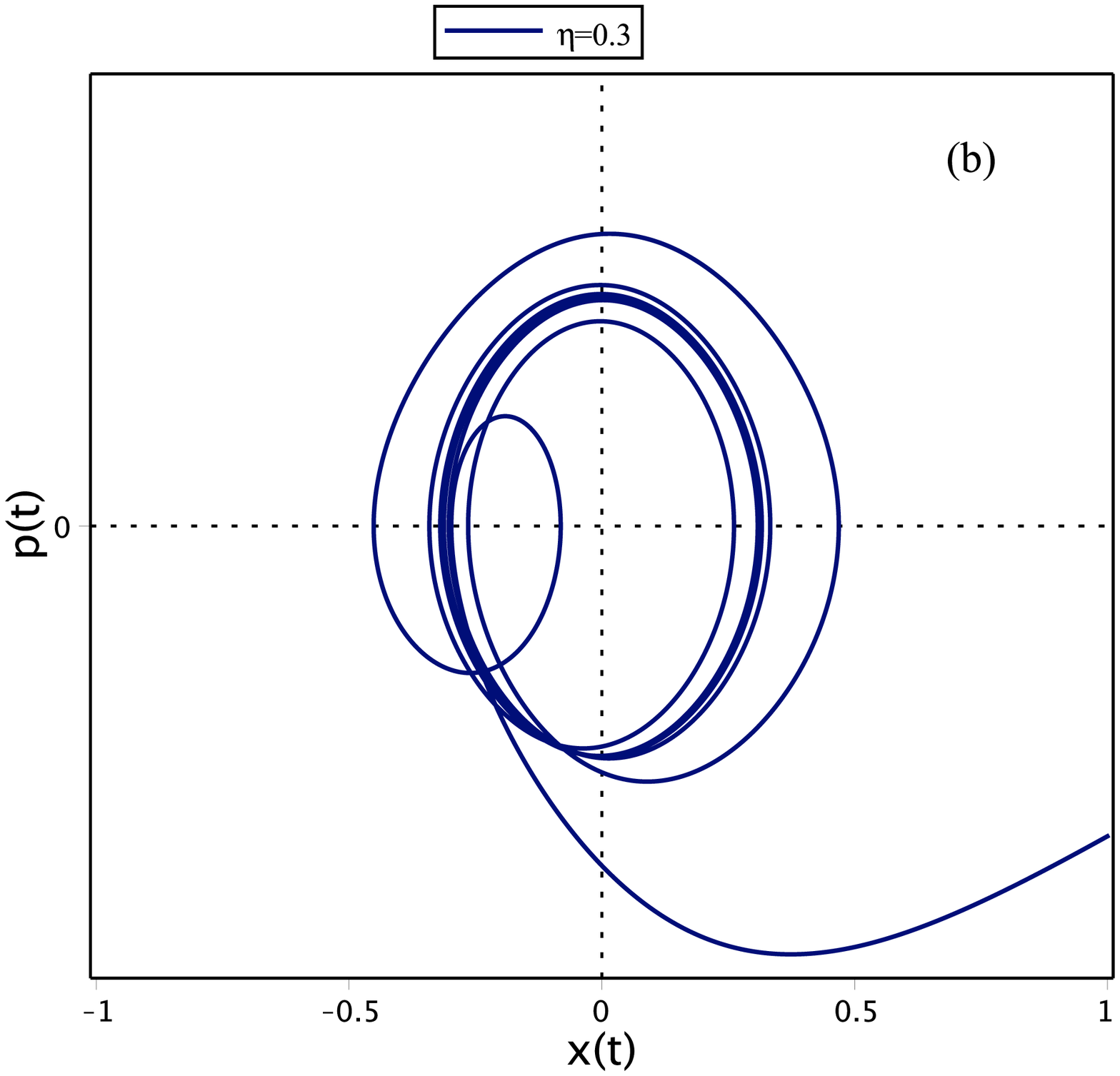} 
\includegraphics[width=0.3\textwidth]{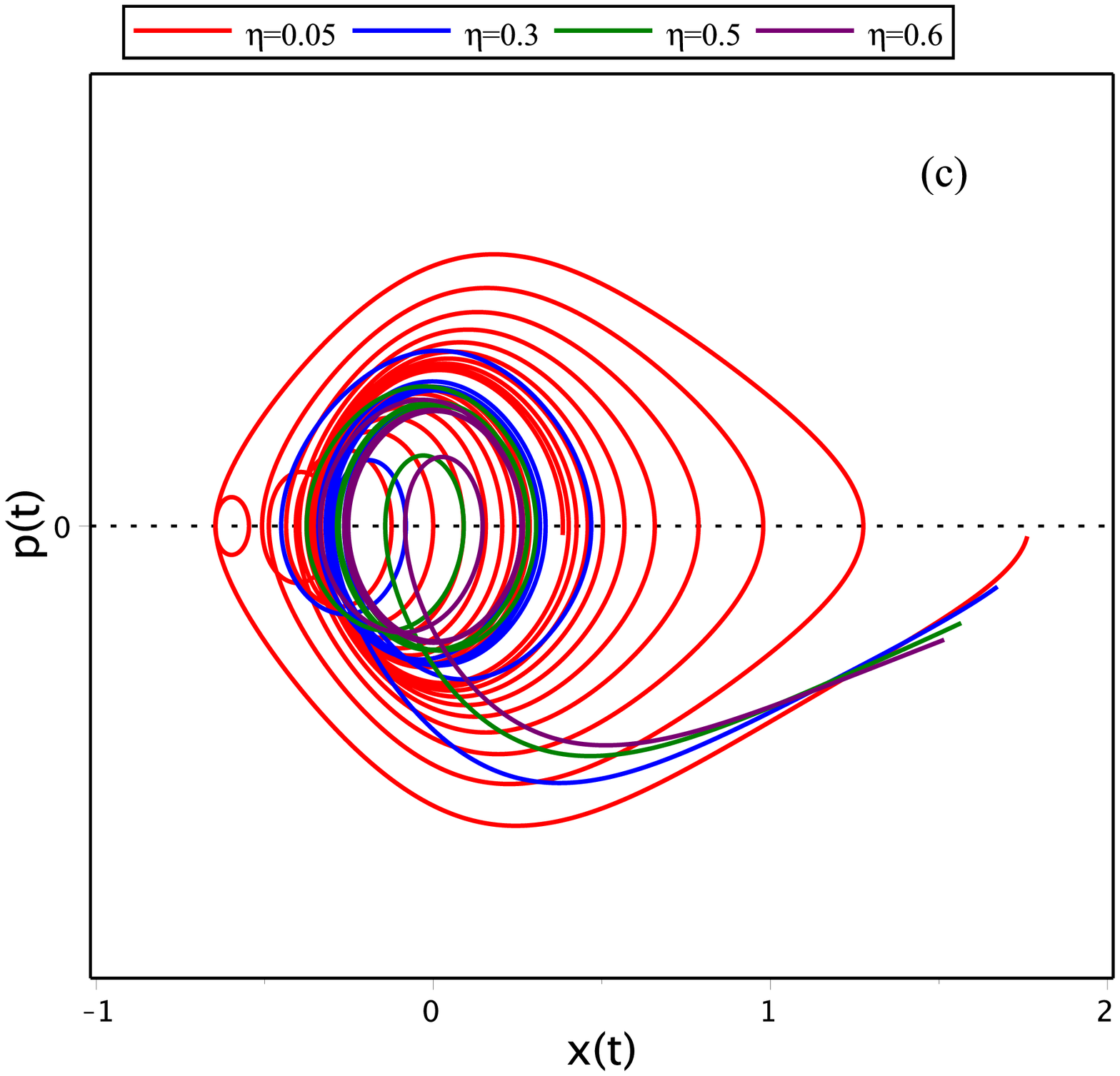}
\caption{\small 
{ For $A_{t}=\lambda=F_{\circ}=\Omega=1$ and $\omega=2$, we show (a) $x(t)$ of (\ref{x(t)-PDM-DDO}) as it evolves in time for different values of the damping parameter $\eta$, (b) the corresponding phase-space trajectory for $\eta=0.3$, and (c) the phase-trajectories for different values of $\eta$.}}
\label{fig1}
\end{figure}%
\begin{figure}[!ht]  
\centering
\includegraphics[width=0.3\textwidth]{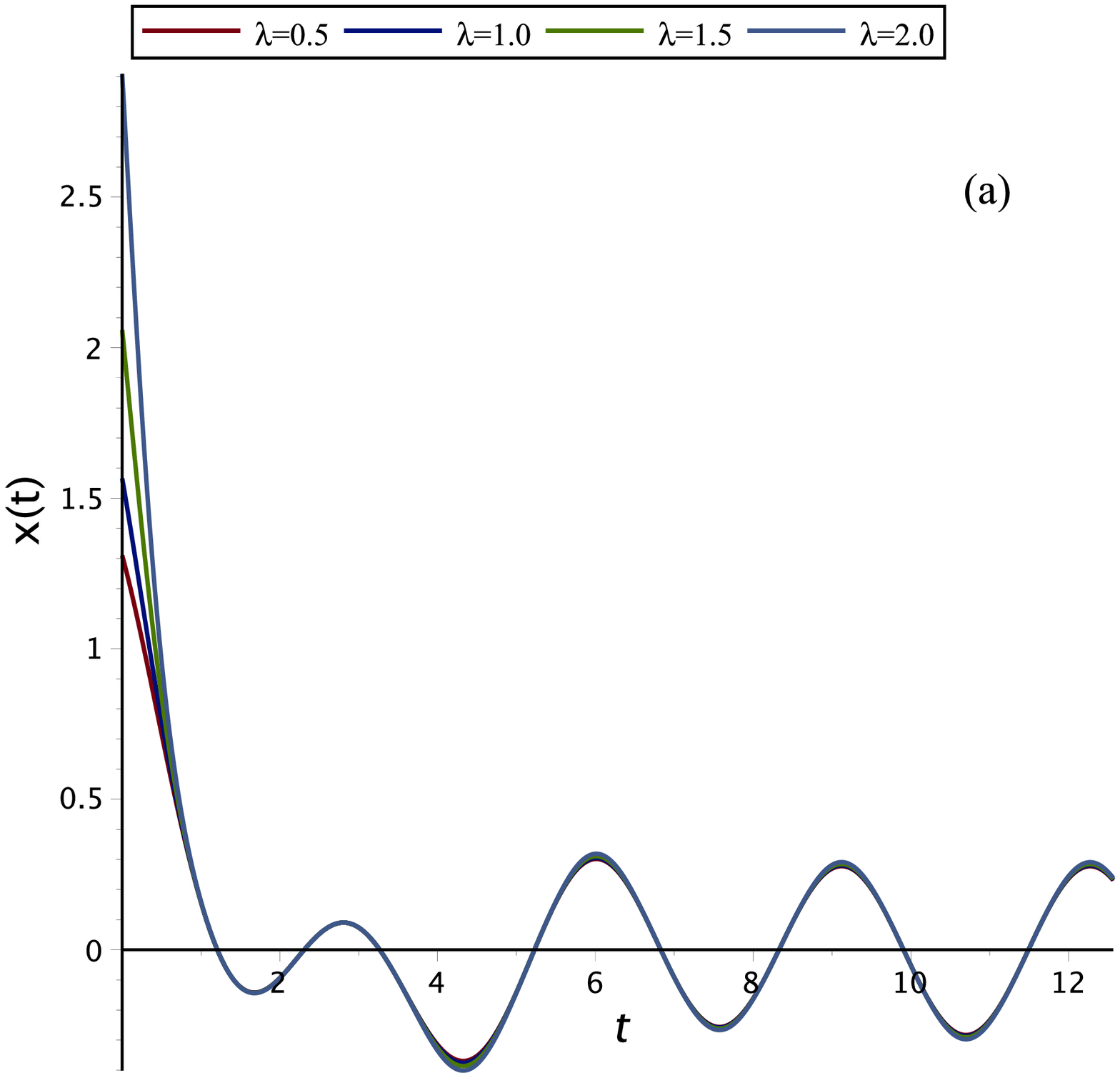}
\includegraphics[width=0.3\textwidth]{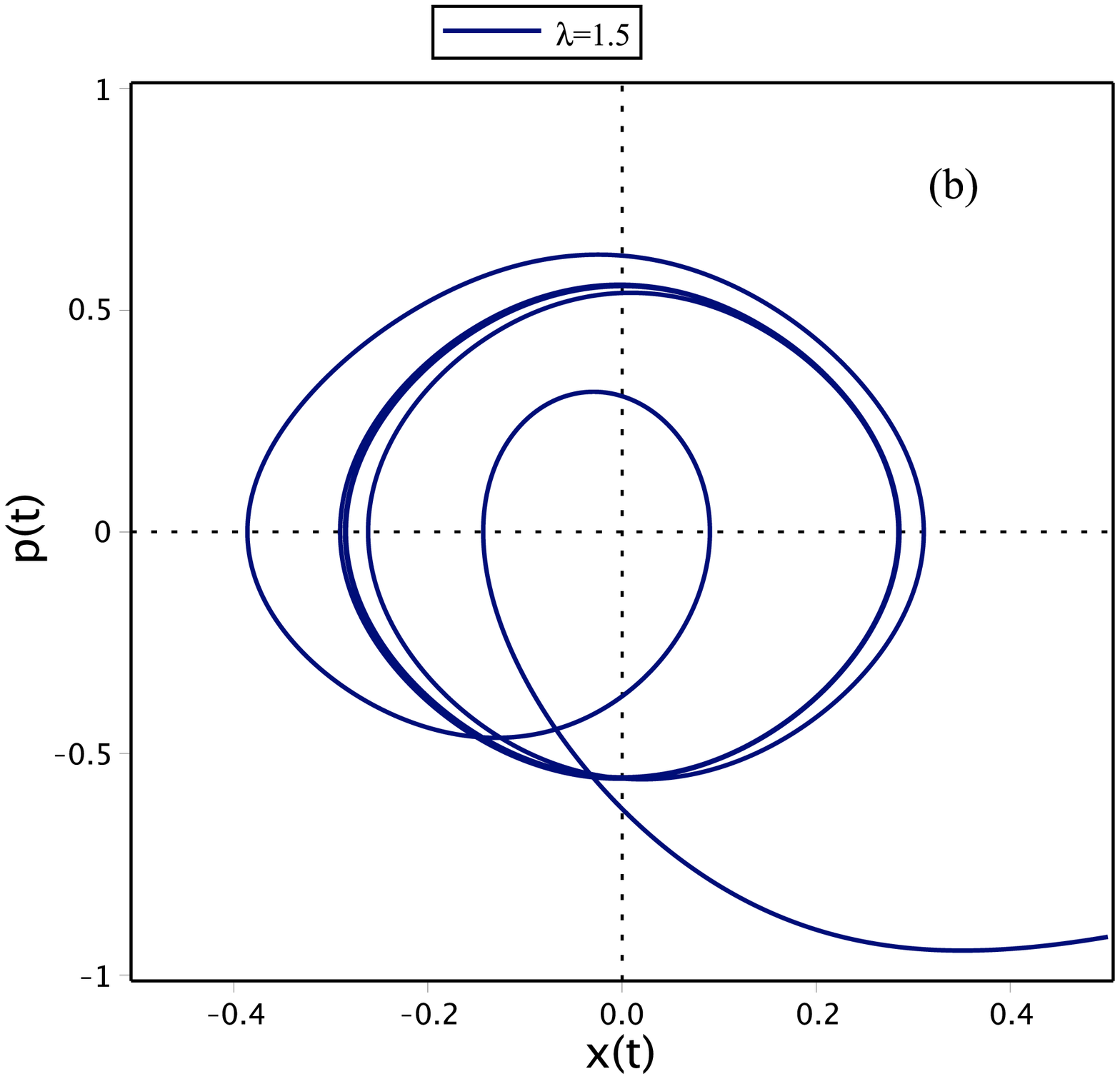} 
\includegraphics[width=0.3\textwidth]{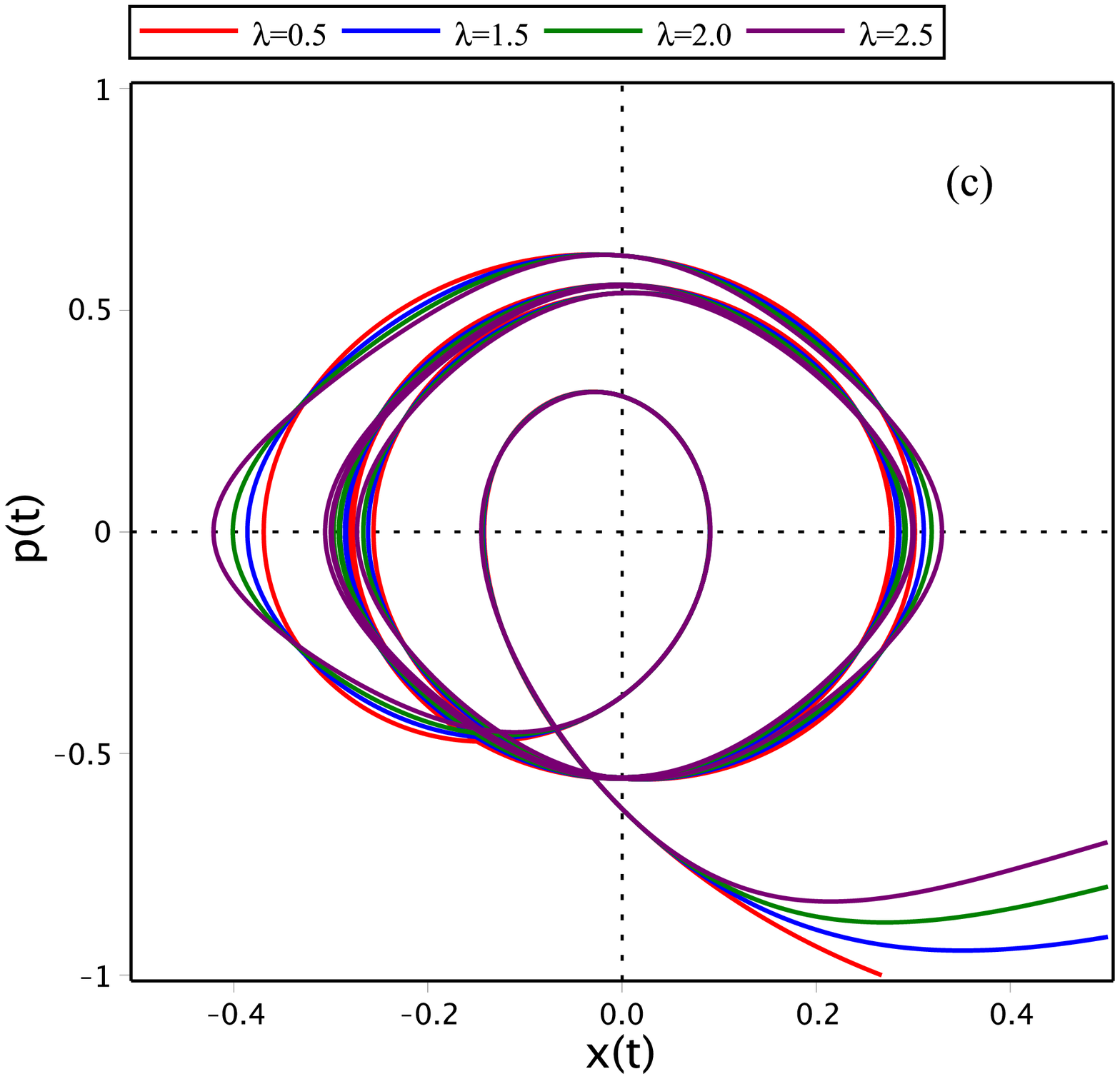}
\caption{\small 
{ For $\eta=0.5, A_{t}=F_{\circ}=\Omega=1$ and $\omega=2$, we show (a) $x(t)$ of (\ref{x(t)-PDM-DDO}) as it evolves in time for different values of the PDM coupling parameter $\lambda$, (b) the corresponding phase-space trajectory for $\lambda=1.5$, and (c) the phase-trajectories for different values of $\lambda$.}}
\label{fig2}
\end{figure}%
Then, the corresponding dynamical equation (\ref{q-DD oscillator eq})
transforms into the PDM-DDO dynamical equation%
\begin{equation}
\ddot{q}\left( t\right) +2\mathcal{\eta }\omega \,\dot{q}\left( t\right)
+\omega ^{2}q\left( t\right) =F_{\circ }\,\cos \left( \Omega t\right)
\Longleftrightarrow \ddot{x}+\frac{m^{\prime }\left( x\right) }{2m\left(
x\right) }\dot{x}^{2}+2\eta \omega \dot{x}+\omega ^{2}\sqrt{\frac{Q\left(
x\right) }{m\left( x\right) }}x=\frac{F_{\circ }}{m\left( x\right) }\,\cos
\left( \Omega t\right) .  \label{PDM-DDO-eq}
\end{equation}%
and inherits its exact solution from (\ref{q(t)-general solution}) through
the point transformation (\ref{PCT1}). Obviously, such a point
transformation secures invariance between the two dynamical systems, the
constant mass system (i.e., the dynamical equation on the left of (\ref%
{PDM-DDO-eq}) and the PDM one (i.e., the dynamical equation on the right of (%
\ref{PDM-DDO-eq})). This methodical proposal is clarified through the
following illustrative example.%
\begin{figure}[!ht]  
\centering
\includegraphics[width=0.3\textwidth]{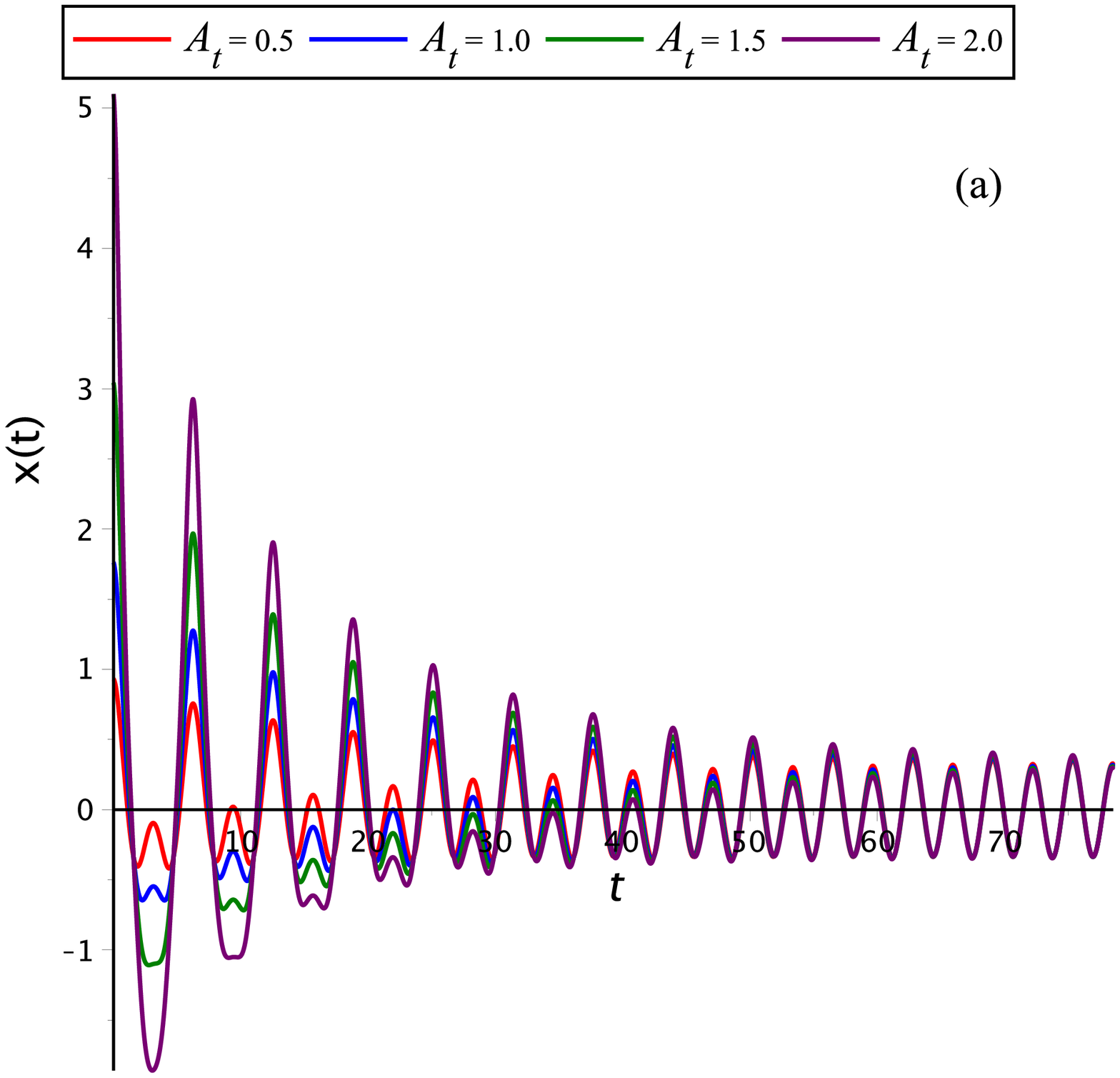}
\includegraphics[width=0.3\textwidth]{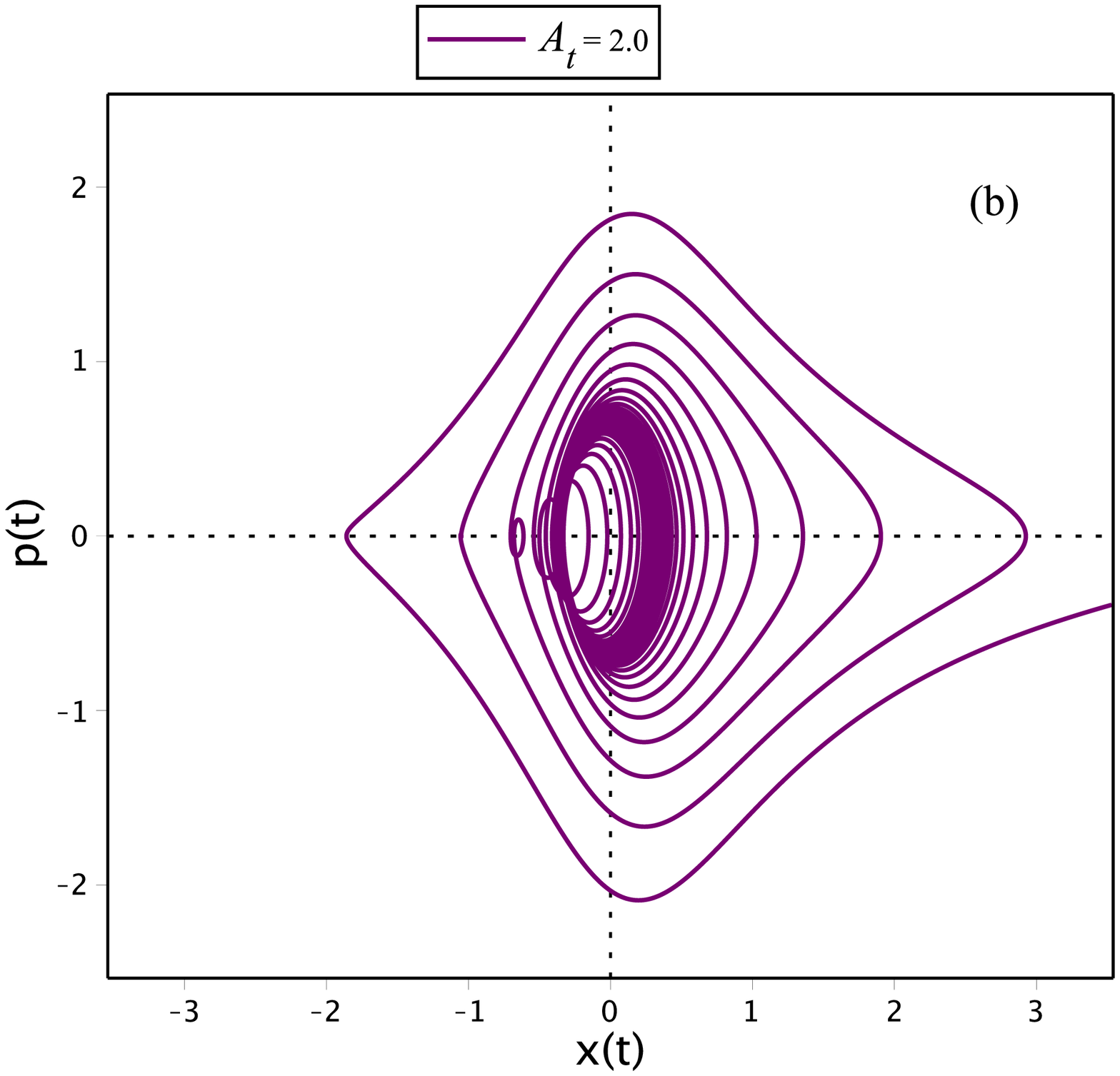} 
\includegraphics[width=0.3\textwidth]{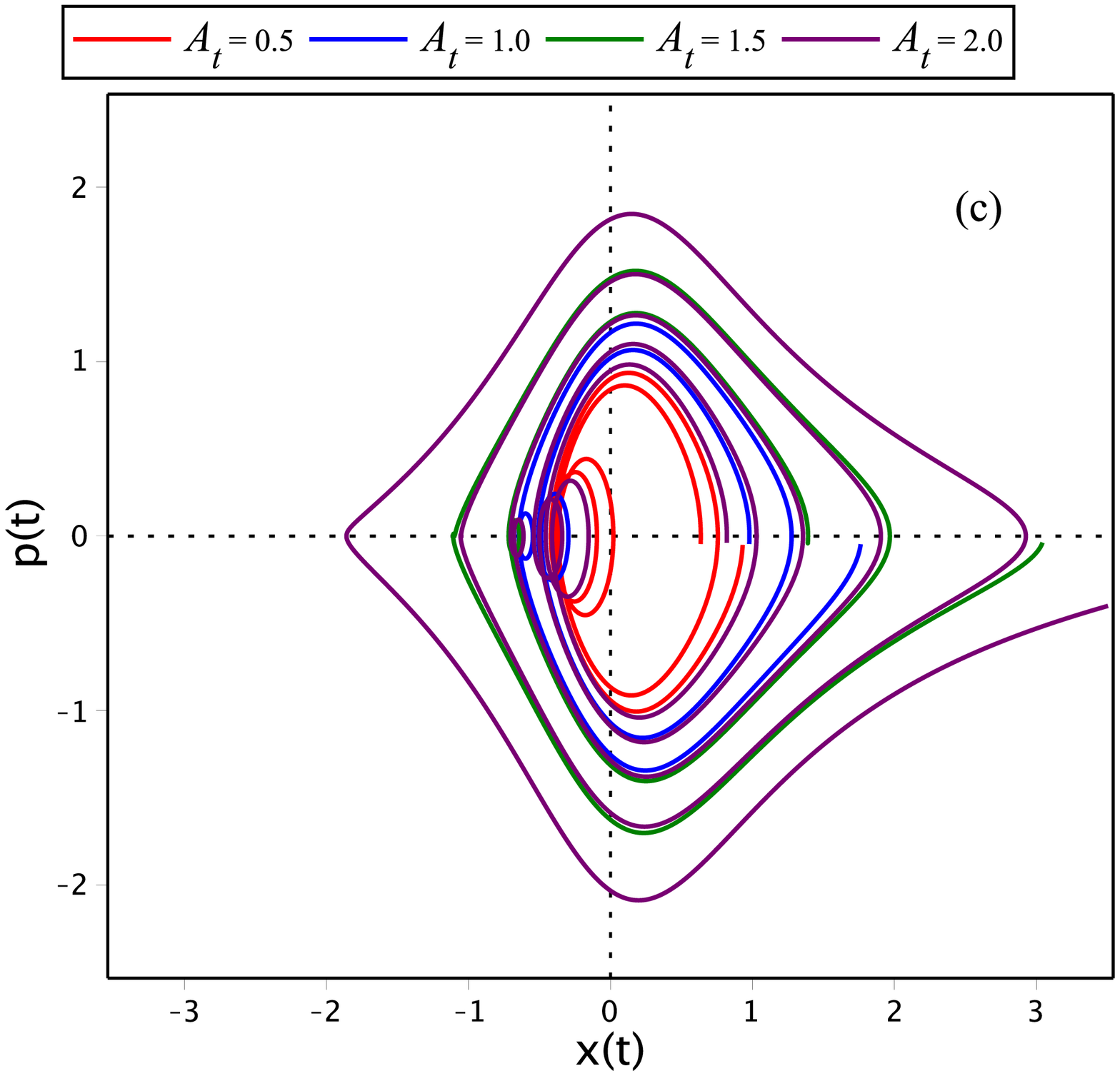}
\caption{\small 
{ For $\eta=0.05$ (under-damping), $\lambda=F_{\circ}=\Omega=1$ and $\omega=2$, we show (a) $x(t)$ of (\ref{x(t)-PDM-DDO}) as it evolves in time for different values of the amplitude of the transient solution $A_{t}$, (b) the corresponding phase-space trajectory for $A_{t}=2$, and (c) the phase-trajectories for different values of $A_{t}$.}}
\label{fig3}
\end{figure}%
\begin{figure}[!ht]  
\centering
\includegraphics[width=0.3\textwidth]{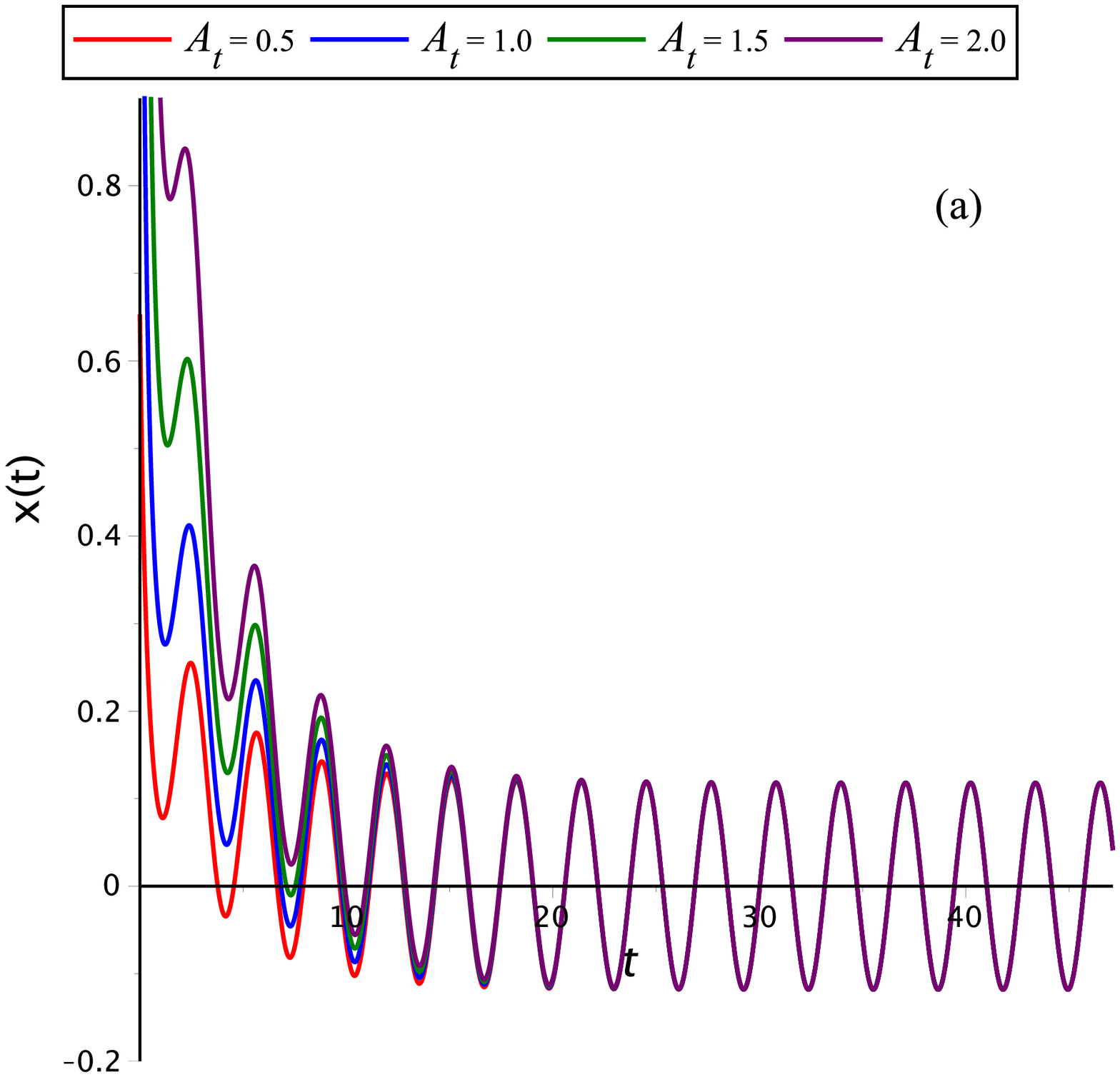}
\includegraphics[width=0.3\textwidth]{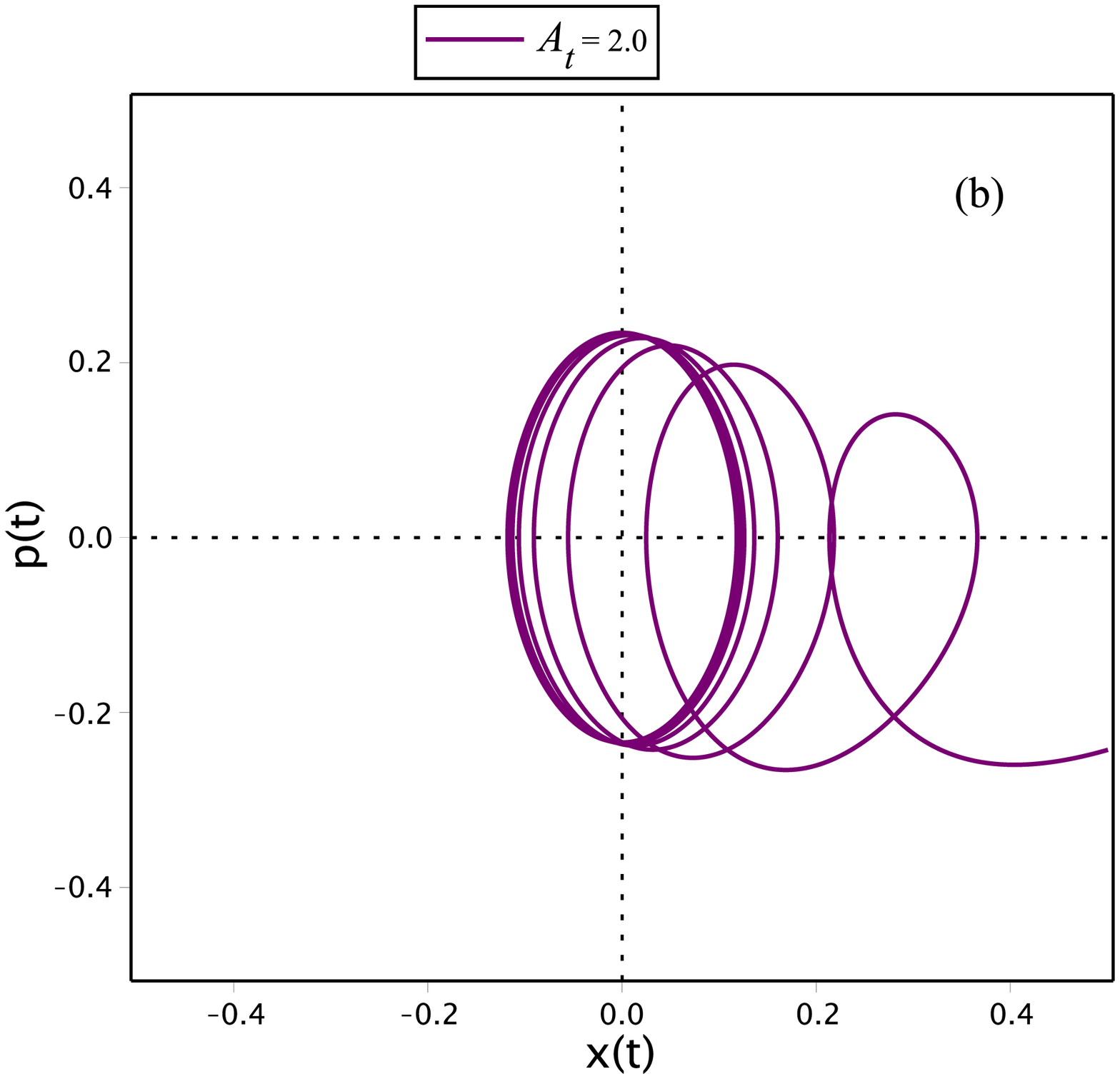} 
\includegraphics[width=0.3\textwidth]{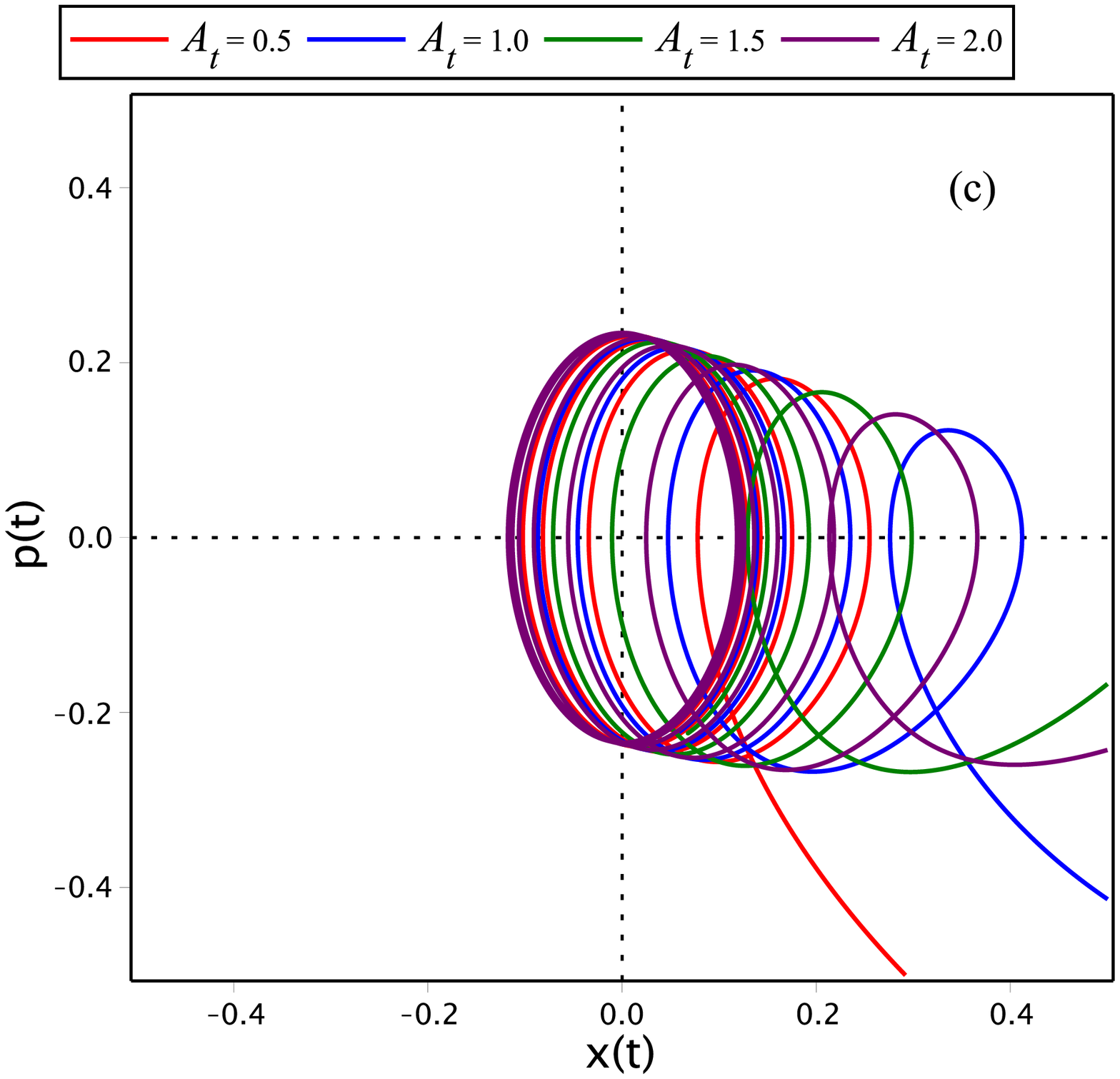}
\caption{\small 
{ For $\eta=2$ (over-damping), $F_{\circ}=\Omega=1$ and $\lambda=\omega=2$, we show (a) $x(t)$ of (\ref{x(t)-PDM-DDO}) as it evolves in time for different values of the amplitude of the transient solution $A_{t}$, (b) the corresponding phase-space trajectory for $A_{t}=2$, and (c) the phase-trajectories for different values of $A_{t}$.}}
\label{fig4}
\end{figure}%

\subsection{A non-singular Mathews-Lakshmanan PDM-DDO model}
Consider a Mathews-Lakshmanan type  \cite{M-L 1974} PDM-particle with 
\begin{equation}
m\left( x\right) =\frac{1}{1+\lambda ^{2}x^{2}}.  \label{ML-PDM-DDO}
\end{equation}%
In this case, our point canonical transformation (\ref{PCT1}) would result in%
\begin{equation}
\sqrt{Q\left( x\right) }x=\frac{1}{\lambda }\ln \left( \lambda x+\sqrt{%
1+\lambda ^{2}x^{2}}\right) .  \label{ML-Q(x)}
\end{equation}%
Using (\ref{q(t)-general solution}) and (\ref{PCT1}), one would write%
\begin{equation}
q\left( t\right) =A_t e^{-\mathcal{\omega }\eta t}\cosh \left( \beta t\right)
+C_s\,\cos \left( \Omega t-\delta \right) =\frac{1}{\lambda }\ln \left(
\lambda x+\sqrt{1+\lambda ^{2}x^{2}}\right) ,  \label{q(t)-x(t)}
\end{equation}%
to imply%
\begin{equation}
x\left( t\right) =\frac{1}{\lambda }\sinh \left( \lambda q\left( t\right)
\right) =\frac{1}{\lambda }\sinh \left( \lambda \left[ A_t e^{-\omega \eta
t}\cosh \left( \beta t\right) +C_s \cos \left( \Omega t-\delta \right) %
\right] \right) ,  \label{x(t)-PDM-DDO}
\end{equation}%
as the exact solution for the PDM-DDO dynamical equation%
\begin{equation}
\ddot{x}-\frac{2\lambda x}{1+\lambda ^{2}x^{2}}\dot{x}^{2}+2\eta \mathcal{%
\omega }\dot{x}+\frac{\mathcal{\omega }}{\lambda }\sqrt{1+\lambda ^{2}x^{2}}%
\ln \left( \lambda x+\sqrt{1+\lambda ^{2}x^{2}}\right) =\left( 1+\lambda
^{2}x^{2}\right) F_{\circ }\,\cos \left( \Omega t\right) .
\label{PDM-DDO eq1}
\end{equation}%
At this point, one should notice that its solution (\ref{x(t)-PDM-DDO})
converges to $q\left( t\right) $ as $\lambda \longrightarrow 0$. That is,%
\begin{equation*}
x\left( t\right) =\lim\limits_{\lambda \longrightarrow 0}\frac{1}{\lambda }%
\sinh \left( \lambda q\left( t\right) \right) =q\left( t\right) .
\end{equation*}%
This PDM-DDO equation of motion  (\ref{PDM-DDO eq1}) describes our PDM-particle $m\left( x\right) $ of (\ref{ML-PDM-DDO}) moving in the vicinity of a conservative potential force field 
\begin{equation}
V\left( x\right) =\frac{1}{2}\mathcal{\omega }^{2}\left( \frac{1}{\lambda }%
\ln \left( \lambda x+\sqrt{1+\lambda ^{2}x^{2}}\right) \right) ^{2},
\label{DDO-V(x)}
\end{equation}%
and feels a non-conservative Rayleigh dissipative force field%
\begin{equation}
\mathcal{R}\left( x,\dot{x}\right) =\frac{b}{2\left( 1+\lambda
^{2}x^{2}\right) }\dot{x}^{2},  \label{DDO-R(x)}
\end{equation}%
along with a driving force $F\left( t\right) =F_{\circ }\,\cos \left( \Omega
t\right) $. Moreover, the corresponding PDM-Lagrangian and PDM-Hamiltonian are given, respectively, by%
\begin{equation}
L\left( x,\dot{x};t\right) =\frac{1}{2}\frac{\dot{x}^{2}}{1+\lambda ^{2}x^{2}%
}-\frac{1}{2}\mathcal{\omega }^{2}\left( \frac{1}{\lambda }\ln \left(
\lambda x+\sqrt{1+\lambda ^{2}x^{2}}\right) \right) ^{2},  \label{PDM-DDO-L1}
\end{equation}%
and%
\begin{equation}
H\left( x,p_{_{x}},t\right) =\left( 1+\lambda ^{2}x^{2}\right) \frac{%
p_{_{x}}^{2}}{2}+\frac{1}{2}\mathcal{\omega }^{2}\left( \frac{1}{\lambda }%
\ln \left( \lambda x+\sqrt{1+\lambda ^{2}x^{2}}\right) \right) ^{2},
\label{PDM-DDO-H1}
\end{equation}%
where%
\begin{equation}
p_{_{x}}=\frac{\partial }{\partial \dot{x}}L\left( x,\dot{x};t\right) =\frac{%
\dot{x}}{1+\lambda ^{2}x^{2}},  \label{PDM-momentum}
\end{equation}%
is the PDM canonical momentum.

From figures 1-4, we clearly observe the competition between the driving and the damping forces to dominate not only the evolution of $x(t)$  of  (\ref{x(t)-PDM-DDO}) but also the evolution of the phase-space trajectories (i.e., evolution of the classical states $\{x(t),p(t)\}$). For example, in figures 1(a), 2(a), 3(a), and 4(a) we observe that the initial domination of the damping force starts to disappear as $x(t)$ evolves in time, where a state of steady oscillatory stability (i.e., the oscillations' amplitudes stabilise as they evolve in time) is achieved as a result of the domination of the driving force at latter time. In figures 1(b), 2(b), 3(b), and 4(b), we see that the phase-space trajectories start to shrink (as is the case with the damped harmonic oscillators ( e.g., as observed in ref. \cite{Mustafa1 2021}) and expand until they stabilize (dark colored lines) into an almost simple harmonic oscillator ones. The same things happen in figures 1(c), 2(c), 3(c), and 4(c) at different parametric settings. Therein, we observe that for each parametric value we find a corresponding stabilization into almost harmonic oscillator phase-space trajectory (dark colored lines).  Moreover, the phase-space trajectories crossing (i.e., classical states $\{x(t),p(t)\}$ crossing) is observed and documented in figures 1(c), 2(c), 3(c), and 4(c) at different parametric values. Yet for a specific parametric value, a state may have multiple crossing with itself as documented in figures 1(b), 2(b), 3(b), and 4(b). 

\subsection{A power-law PDM-DDO model}
A power-law type coordinate transformation%
\begin{figure}[!ht]  
\centering
\includegraphics[width=0.3\textwidth]{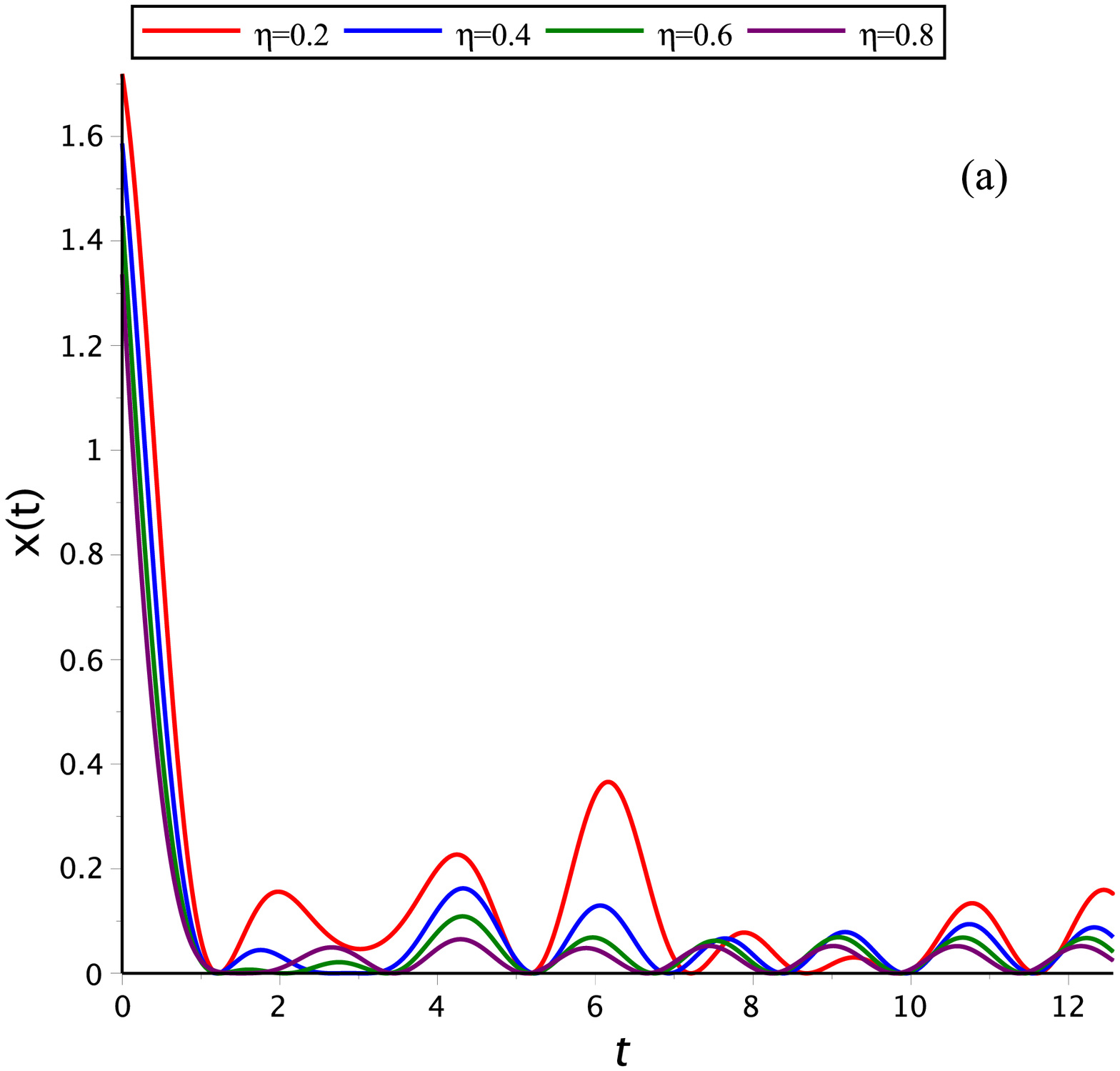}
\includegraphics[width=0.3\textwidth]{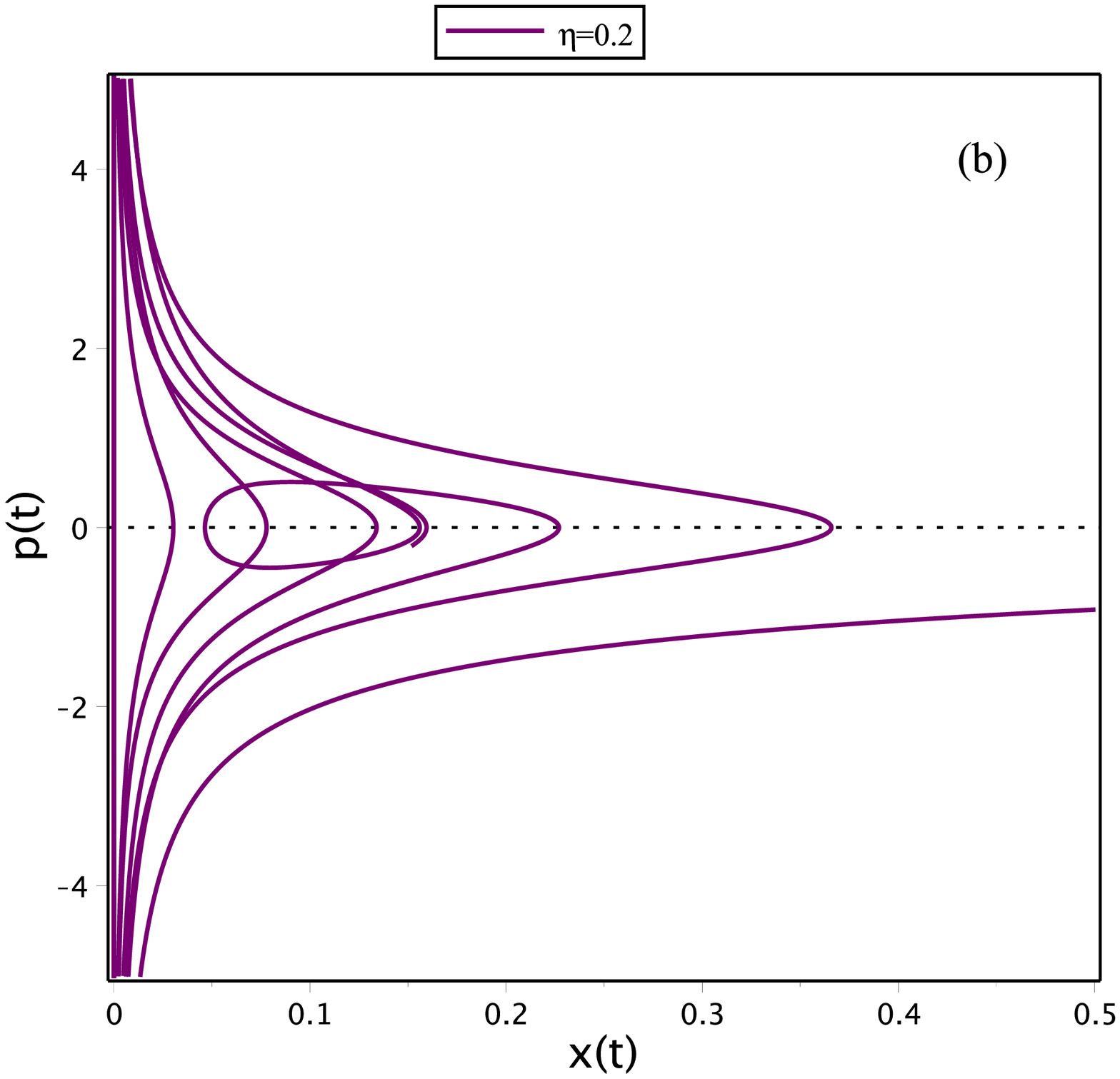} 
\includegraphics[width=0.3\textwidth]{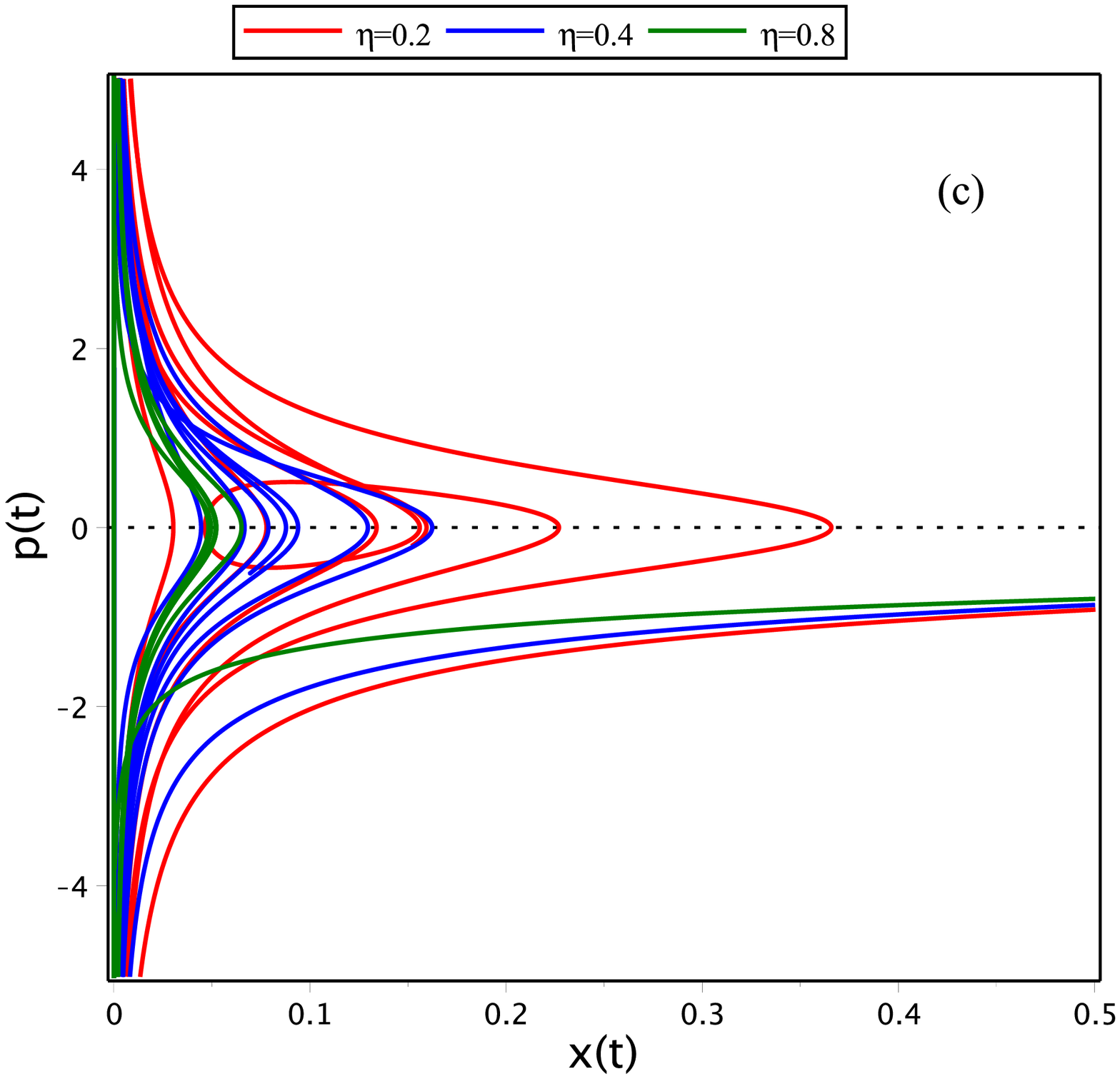}
\caption{\small 
{ For $a=A_t=F_{\circ}=\Omega=1$, $\omega=2$, and $\eta < 1$ (i.e., under-damping) we show (a) $x(t)$ of (29) as it evolves in time, (b) the corresponding phase-space trajectory for $\eta=0.2$, and (c) the phase-trajectories for $\eta=0.2,0.4,0.8$.}}
\label{fig5}
\end{figure}%
\begin{figure}[!ht]  
\centering
\includegraphics[width=0.3\textwidth]{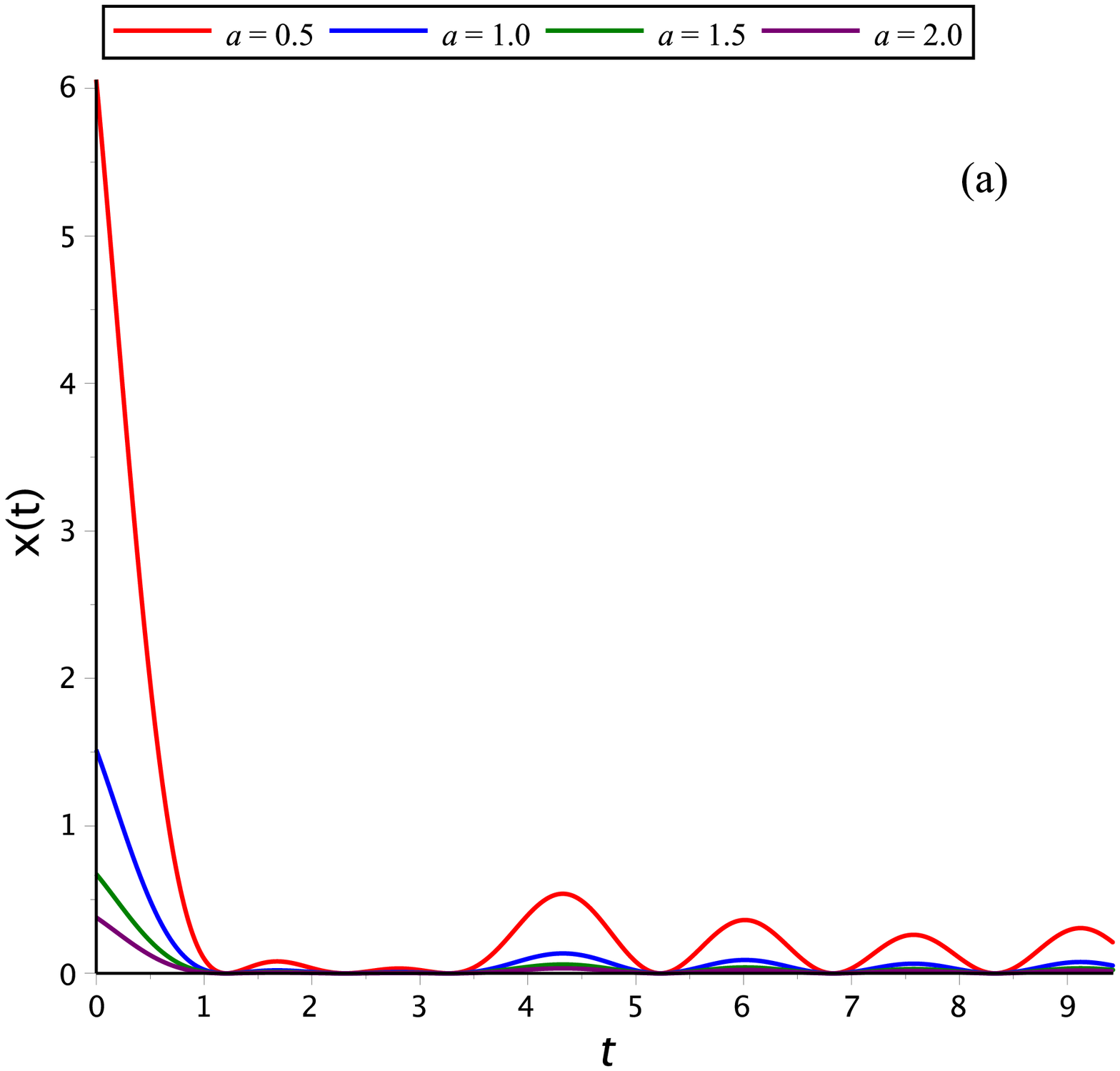}
\includegraphics[width=0.3\textwidth]{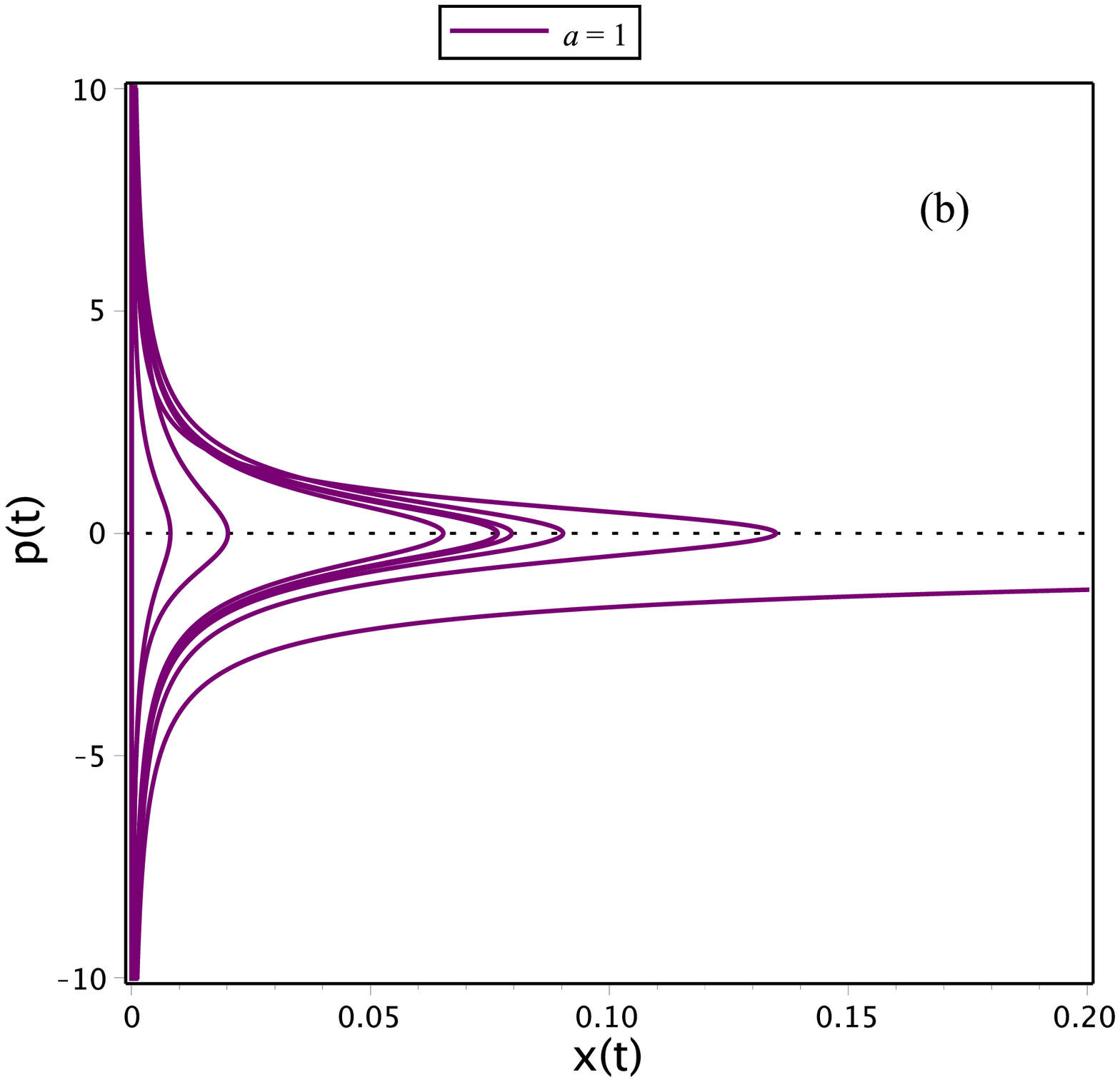} 
\includegraphics[width=0.3\textwidth]{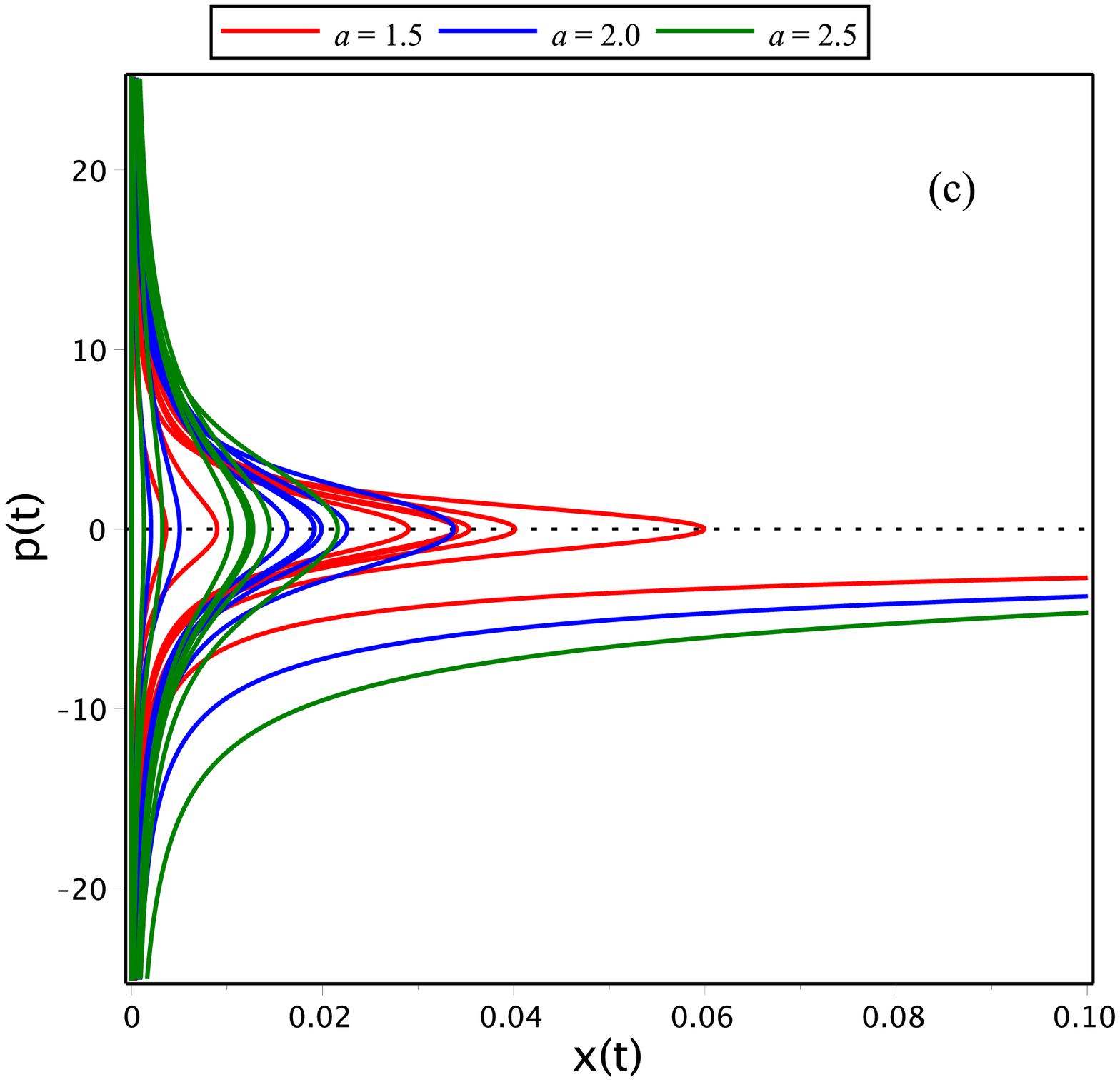}
\caption{\small 
{ For $\eta=0.5$ (under-damping), $A_t=F_{\circ}=\Omega=1$, $\omega=2$, and for different values of the PDM parameter $a$ of (28) we show (a) $x(t)$ of (29) as it evolves in time, (b) the corresponding phase-space trajectory for $a=1$, and (c) the phase-space trajectories for $a=1.5,2,2.5$.}}
\label{fig6}
\end{figure}%
\begin{equation}
\sqrt{Q(x)}=a\,x^\nu,  \label{Q(x) power-law}
\end{equation}%
would imply that%
\begin{equation}
m(x)=a^2\,(\nu+1)^2\,x^{2\nu}; \,  \nu\neq 0,1 .\label{m(x) power-law}
\end{equation}%
In this case, our point canonical transformation (\ref{PCT1}) would result in%
\begin{equation}
x(t)=\left(\frac{q(t)}{a}\right)^{1/(\nu+1)}=\left(\frac{A_t e^{-\mathcal{\omega }\eta t}\cosh \left( \beta t\right)+C_s\,\cos \left( \Omega t-\delta \right)}{a}\right)^{1/(\nu+1)},  \label{x(t) power-law}
\end{equation}%
This is the exact solution for the power-law PDM-DDO dynamical equation%
\begin{equation}
\ddot{x}(t)+\frac{\nu}{x(t)}\dot{x}(t)^2+2\eta\omega\dot{x}(t)+\frac{\omega^2}{\nu+1}x(t)=\frac{F_o\, cos(\Omega t)}{a^2(\nu+1)^2x(t)^{2\nu}}; \, \nu\neq 0,1 .\label{PDM-DDO power-law}
\end{equation}%
Which describes a power-law-type PDM (\ref{m(x) power-law}) moving in a conservative potential force field%
\begin{equation}
V(x)=\frac{a^2\omega^2}{2}x^{2(\nu+1)} \label{V(x) power-law}
\end{equation}%
and feels a non-conservative Rayleigh dissipative force field%
\begin{figure}[!ht]  
\centering
\includegraphics[width=0.3\textwidth]{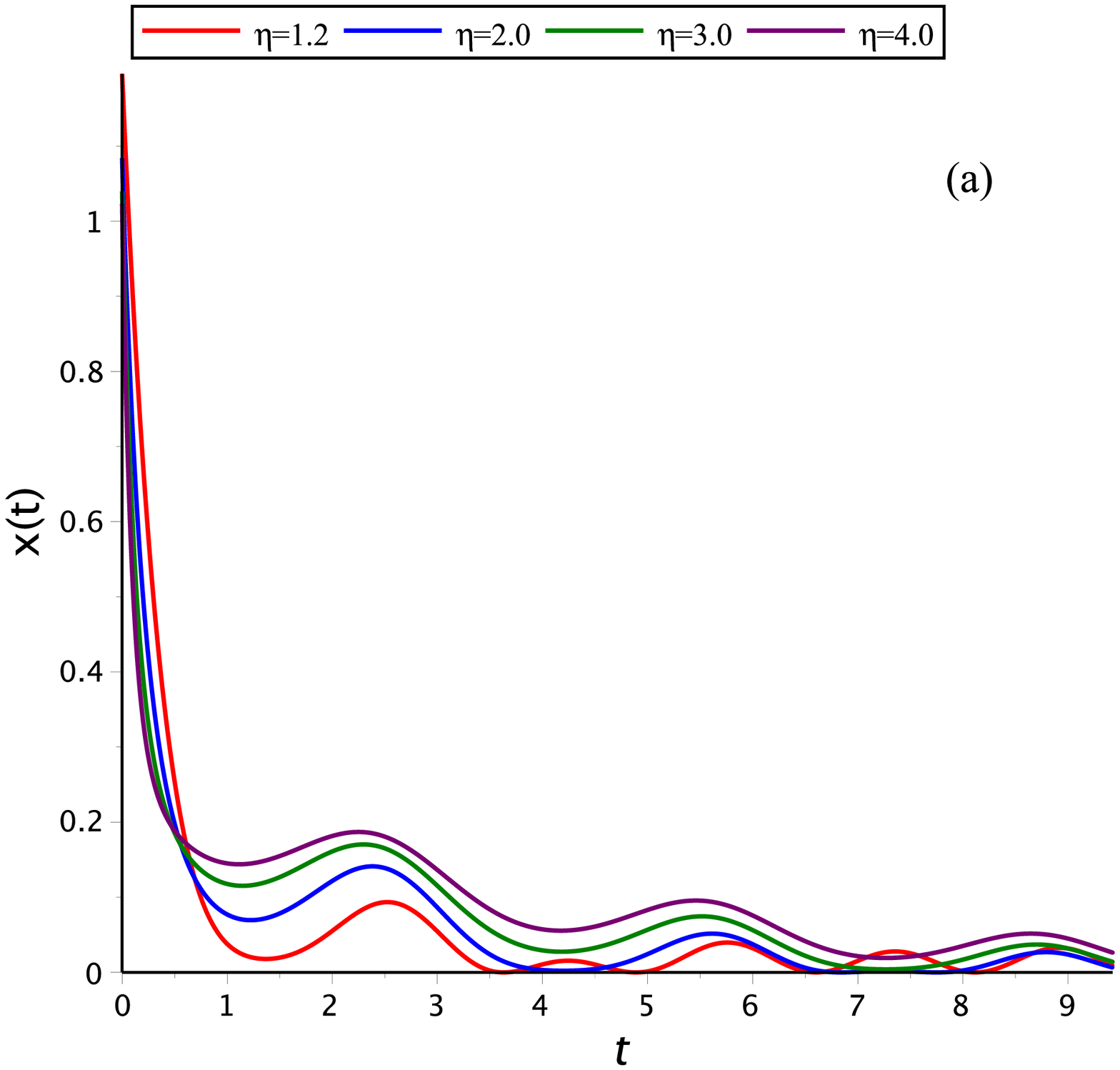}
\includegraphics[width=0.3\textwidth]{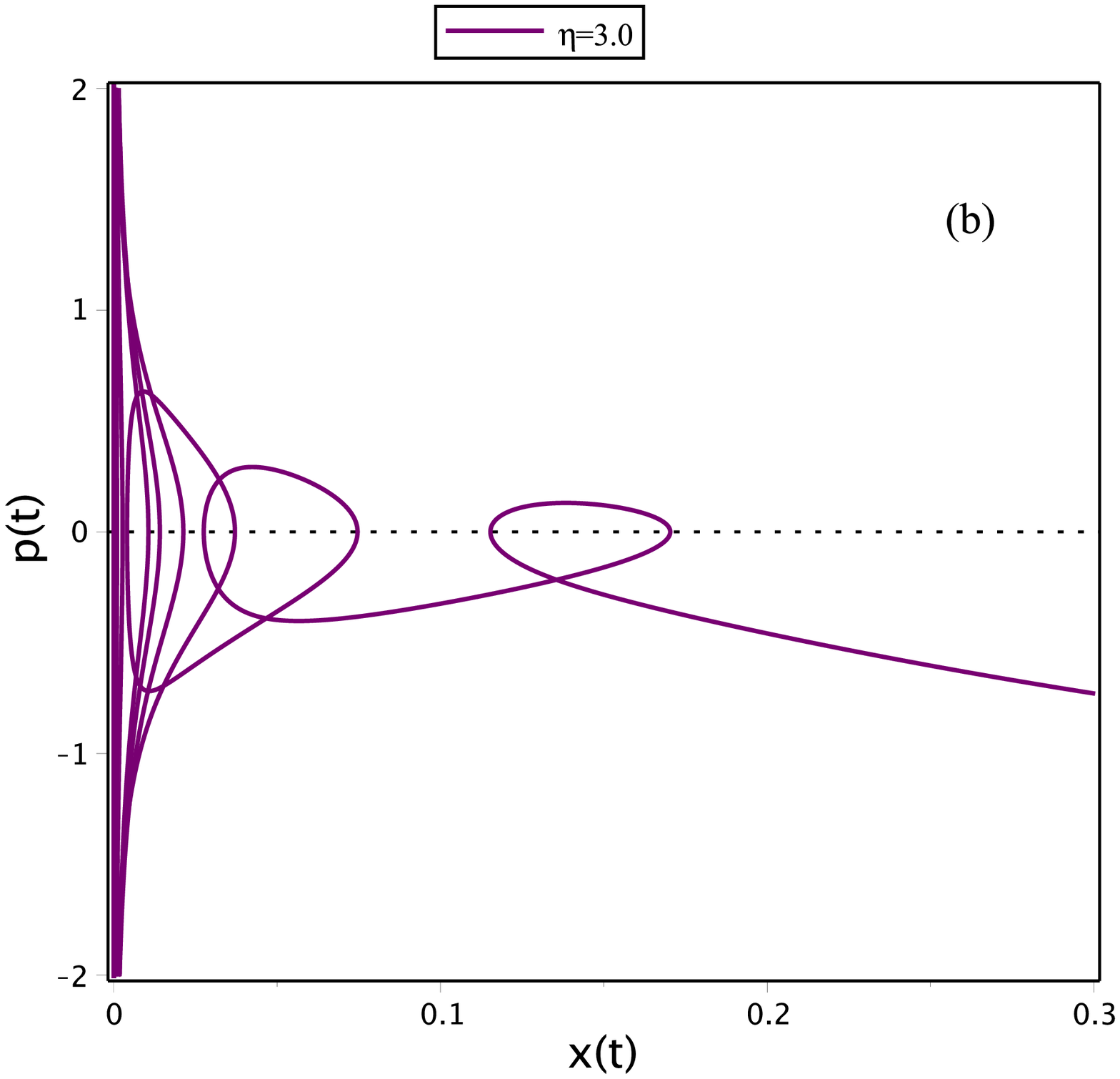} 
\includegraphics[width=0.3\textwidth]{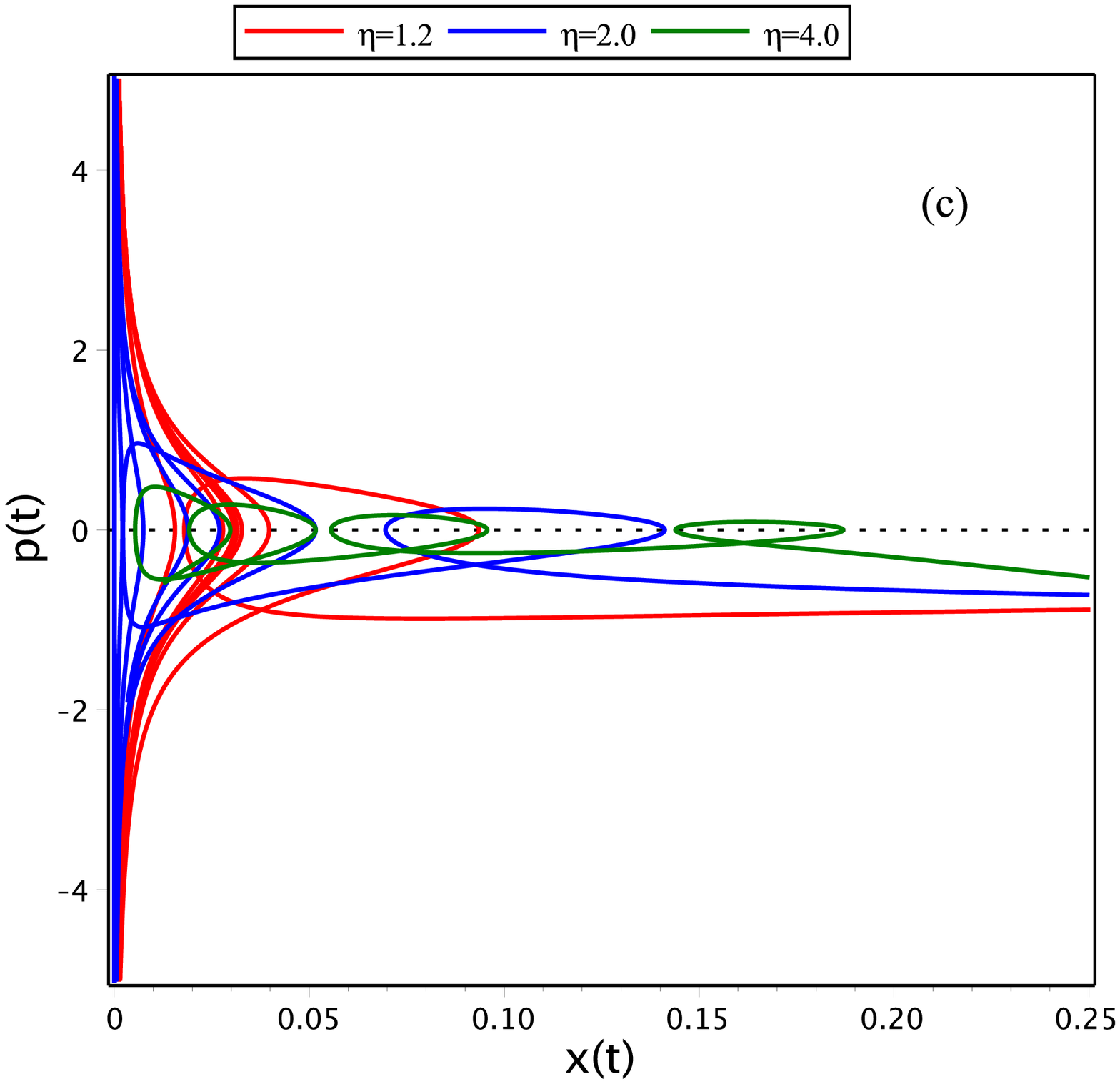}
\caption{\small 
{ For $a=A_t=F_{\circ}=\Omega=1$, $\omega=2$, and $\eta>1$ (over-damping) we show (a) $x(t)$ of (29) as it evolves in time, (b) the corresponding phase-space trajectory for $\eta=3$, and (c) the phase-trajectories for $\eta=1.2,2,4$.}}
\label{fig7}
\end{figure}%
\begin{figure}[!ht]  
\centering
\includegraphics[width=0.3\textwidth]{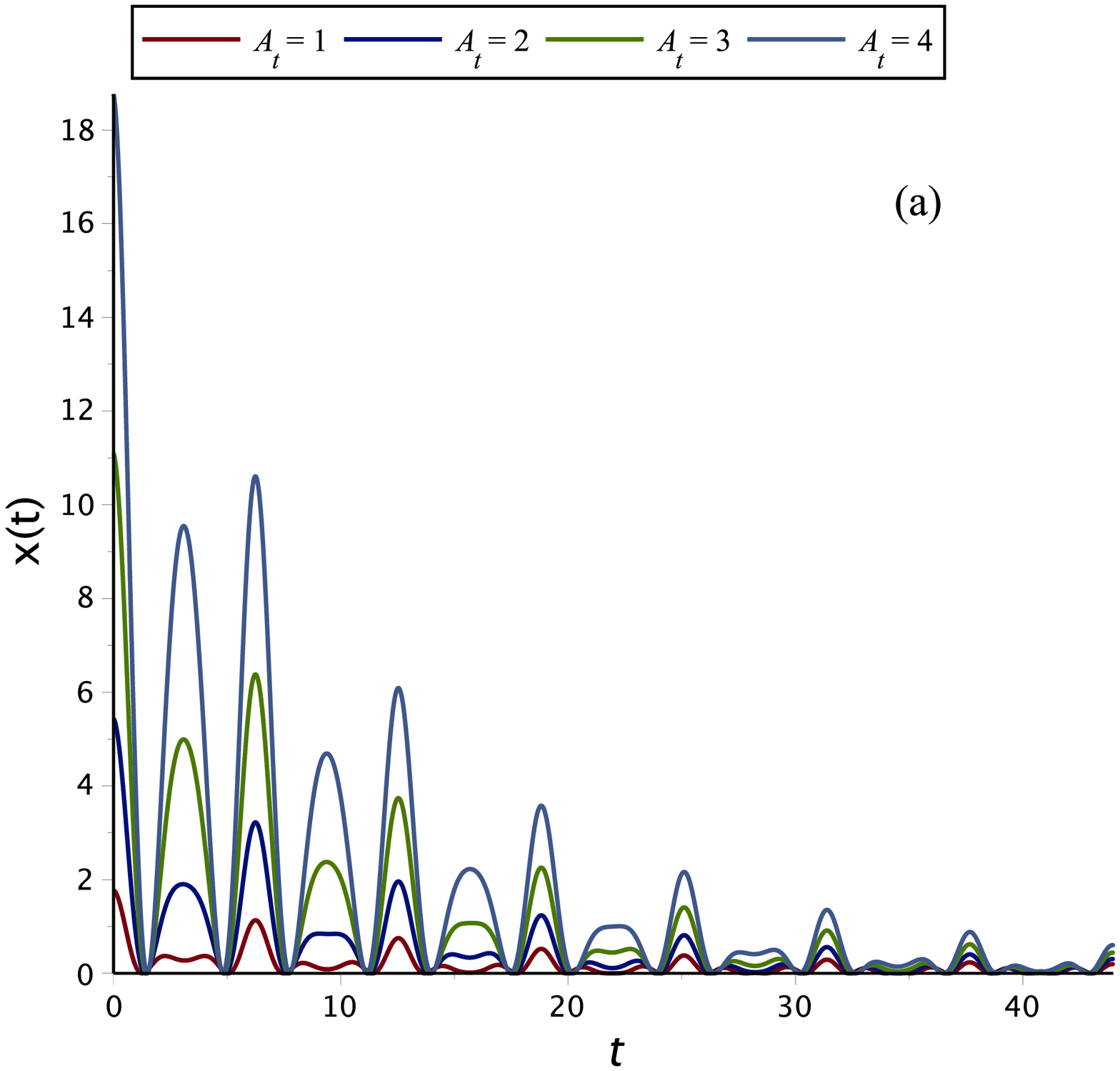}
\includegraphics[width=0.3\textwidth]{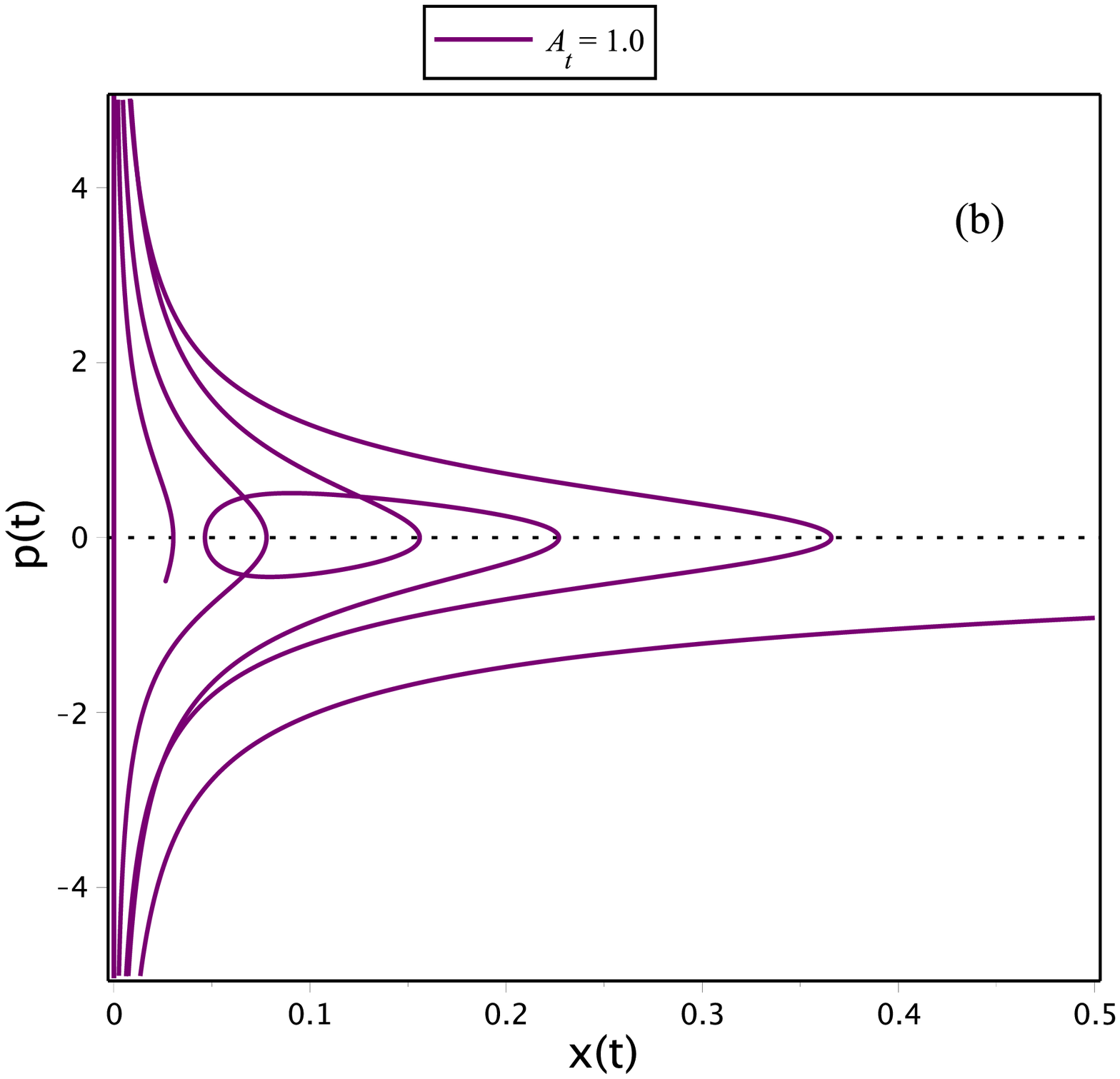} 
\includegraphics[width=0.3\textwidth]{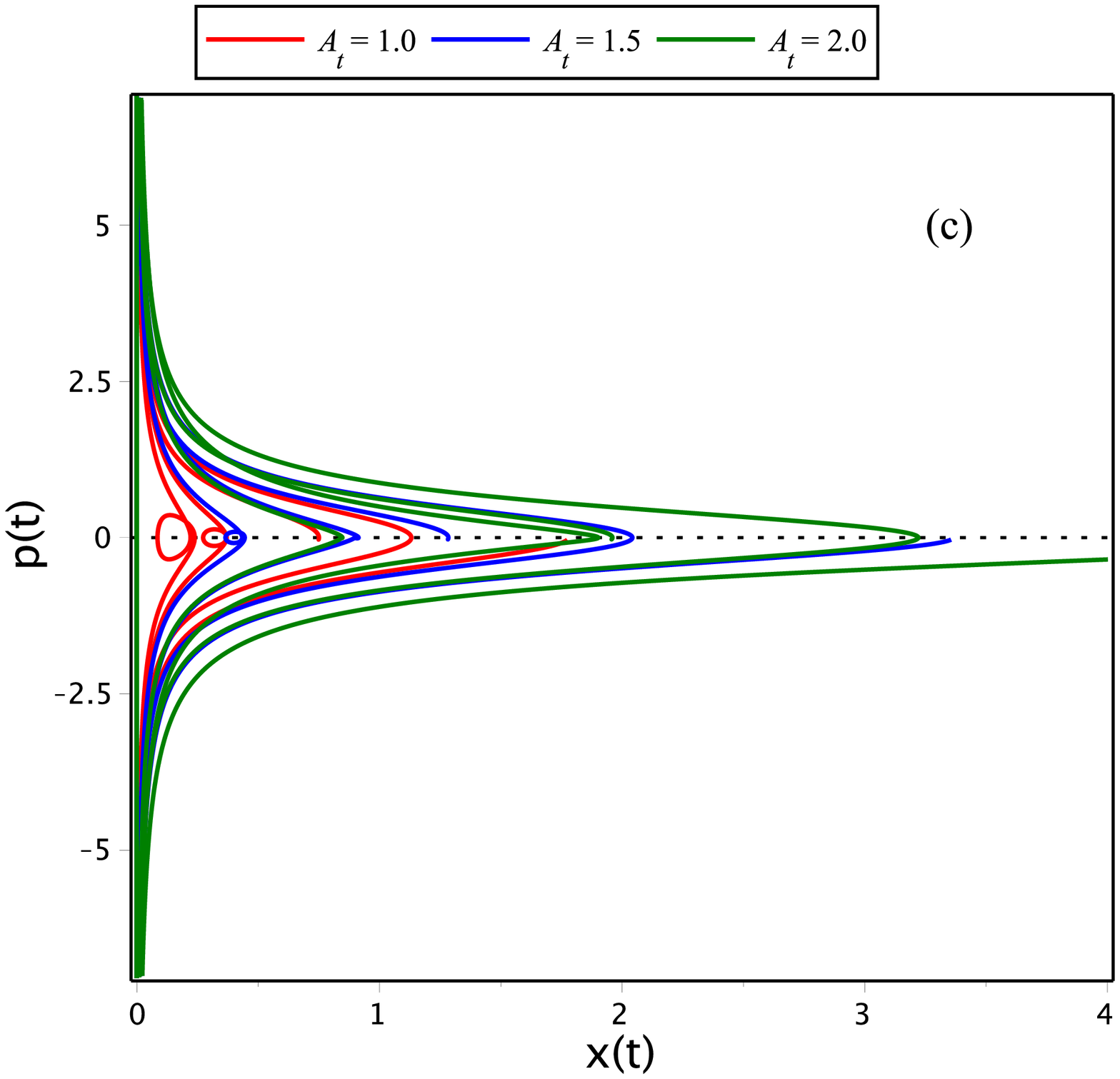}
\caption{\small 
{ For $\eta=0.05$ (under-damping), $a=F_{\circ}=\Omega=1$, $\omega=2$, and for different values of the amplitude of the transient solution $A_{t}$ we show (a) $x(t)$ of (29) as it evolves in time, (b) the corresponding phase-space trajectory for $A_{t}=1$, and (c) the phase-trajectories for $A_{t}=1,1.5,2$.}}
\label{fig8}
\end{figure}%
\begin{equation}
\mathcal{R}\left( x,\dot{x}\right) =\frac{b}{2}\,a^2(\nu+1)^2x^{2\nu}\dot{x}^{2}.  \label{power-law DDO-R(x)}
\end{equation}%
The corresponding PDM Lagrangian and PDM Hamiltonian are, respectively, given by%
\begin{equation}
L\left( x,\dot{x};t\right) =\frac{1}{2}a^2(\nu+1)^2x^{2\nu}\dot{x}^{2}-\frac{a^2\omega^2}{2}x^{2(\nu+1)} ,  \label{PDM-DDO-L2}
\end{equation}%
and%
\begin{equation}
H\left( x,p_{_{x}},t\right) =\frac{p_{_{x}}^{2}}{2a^2\,(\nu+1)^2\,x^{2\nu}}+\frac{a^2\omega^2}{2}x^{2(\nu+1)},
\label{PDM-DDO-H2}
\end{equation}%
where%
\begin{equation}
p_{_{x}}=\frac{\partial }{\partial \dot{x}}L\left( x,\dot{x};t\right) =a^2\,(\nu+1)^2\,x^{2\nu}\dot{x},  \label{PDM-DDO momentum}
\end{equation}%
\begin{figure}[!ht]  
\centering
\includegraphics[width=0.3\textwidth]{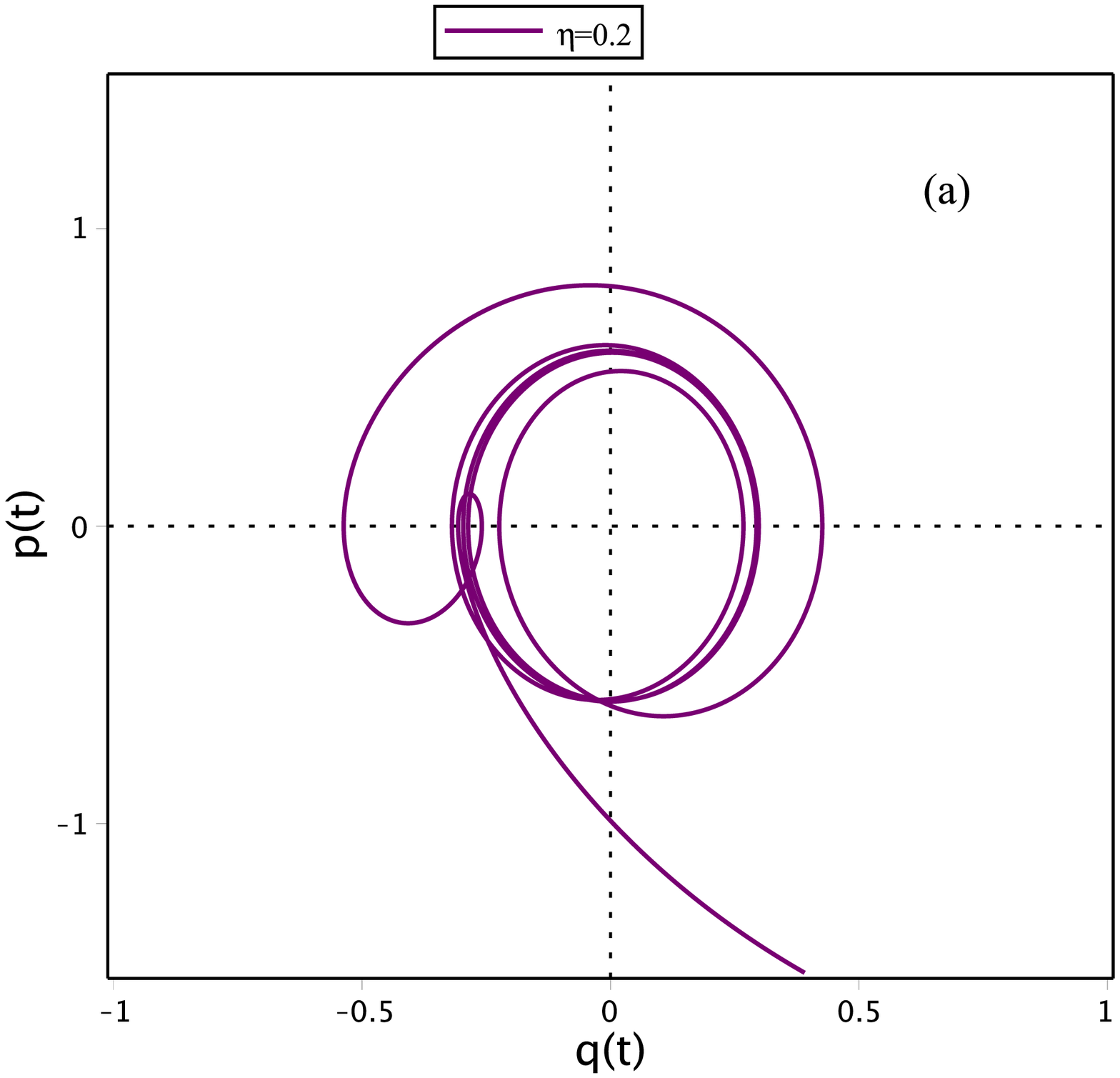}
\includegraphics[width=0.3\textwidth]{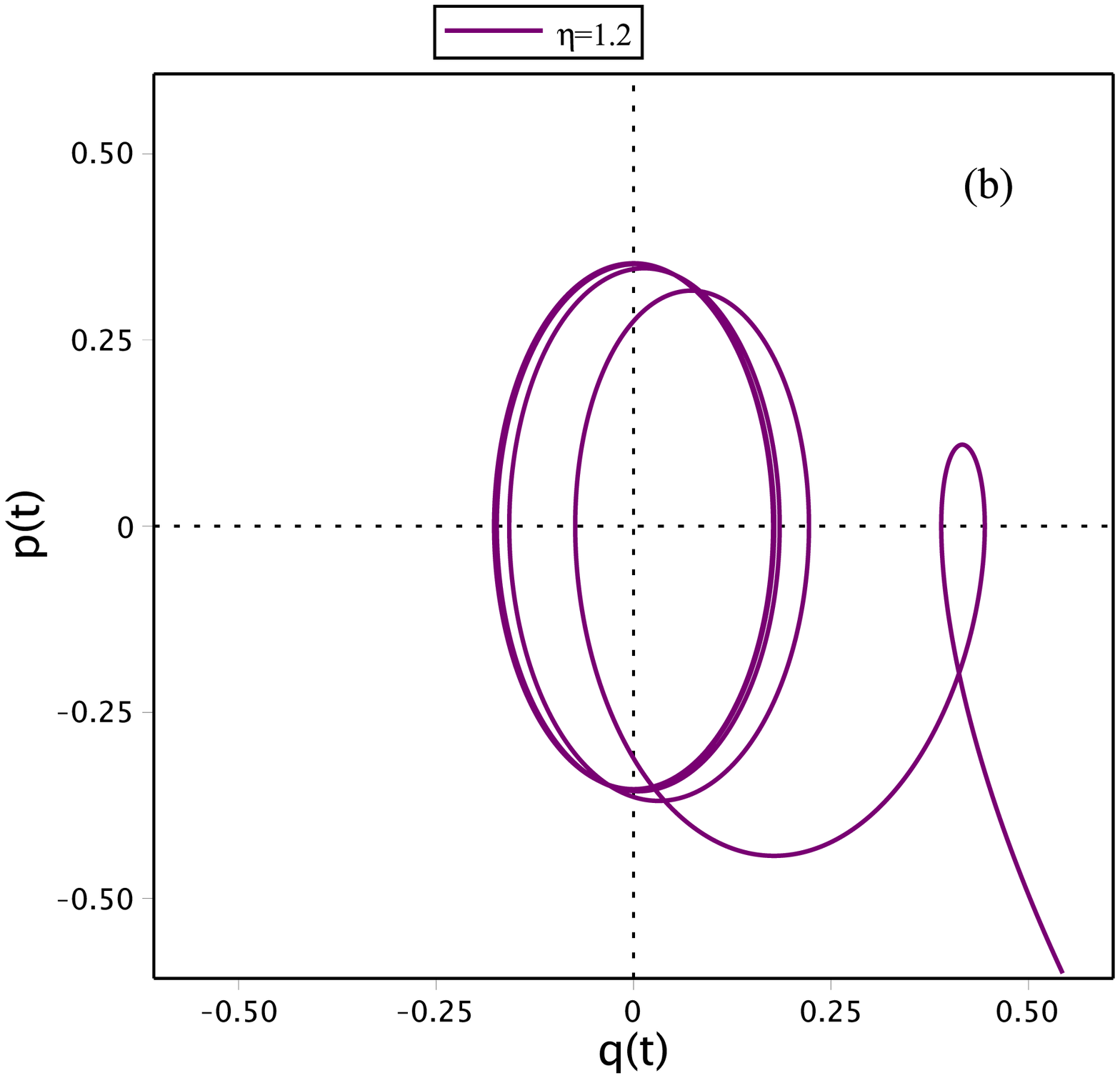} 
\caption{\small 
{ For $A_t=2$, $F_{\circ}=\Omega=1$, $\omega=2$, and for different values of $\eta$ we show the phase-space trajectories for a classical particle with constant mass $m_{\circ}=1$ settings of $q(t)$ in (\ref{q(t)-general solution}) and its corresponding canonical momentum $p(t)$ (a) at $\eta=0.2)$ (under-damping), and (b) at $\eta=1.2$ (over-damping) as the states evolve in time.}}
\label{fig9}
\end{figure}%
is the PDM canonical momentum. 

Figures 5-8, once again show the competition between the driving and the damping forces is clearly observed. In figures 5(a), 6(a), 7(a), and 8(a) we observe that the initial domination of the damping force dies out as $x(t)$ evolves in time, where the oscillations' amplitudes stabilise as a consequence of the domination of the driving force at latter time. In figures 5(b), 6(b), 7(b), and 8(b), we see that the phase-space trajectories start to shrink and expand until they stabilize (dark colored lines). The same trend repeats itself in figures 5(c), 6(c), 7(c), and 8(c) at different parametric settings. Where, for each parametric value we find a corresponding stabilization of the phase-space trajectory (dark colored lines).  Once more, classical states $\{x(t),p(t)\}$ crossing is observed and documented. Yet for a specific parametric value, a state may have multiple crossing with itself as documented in figures 5(b), 6(b), 7(b), and 8(b). 

For both PDM illustrative examples above, Such a behavior trend is also observed for classical particles with constant mass settings. This is due to the first exponentially decay term in the argument of $sinh$ function of (\ref{x(t)-PDM-DDO}), where its contribution dies out and the second term in (\ref{x(t)-PDM-DDO}) dominants over at latter time, as in Figure 9.%

\section{Concluding Remarks}

To make our methodical proposal self-contained, we have recollected the mathematical preliminaries for a constant mass $m_{\circ}=1$ classical particle performing a DDO-motion under the influence of a conservative harmonic oscillator force field $V\left( q\right) =\frac{1}{2}\omega ^{2}q^{2}$, a Rayleigh dissipative force field $\mathcal{R}\left( \dot{q}\right) =\frac{1%
}{2}b\,\dot{q}^{2}$, and in the presence of an external periodic non-autonomous force $F\left( t\right) =F_{\circ }\,\cos \left( \Omega t\right) $. We have used the point canonical transformation (\ref{PCT1})  and reported the corresponding  PDM-Lagrangian (\ref{PDM-DDO-L}) as well as the corresponding PDM dynamical equation for the DDO (\ref{PDM-DDO-eq}) . We have used two illustrative examples: a non-singular PDM (\ref{ML-PDM-DDO}) and a power-law PDM one (\ref{m(x) power-law}). To the best of our knowledge, such PDM-DDO proposal has never been reported elsewhere.

Early on, we have reported the phase-space trajectories crossing (i.e., classical states $\{x(t),p(t)\}$ crossings) as a phenomenon associated with dissipative forces (c.f., e.g., \cite{Mustafa arXiv1,Mustafa 2021,Mustafa1 2021})
which may as well emerge as a result of the point canonical transformation (\ref{PCT1}) into PDM-settings. In the current study, we witness yet another type of classical states crossing. That is, as a classical state $\{x_{i}(t),p_{i}(t)\}$ evolves in time it may cross itself at an earlier and/or latter time/s following the recipe%
\begin{equation}
\{x_{i}(t_1),p_{i}(t_1)\}=\{x_{i}(t_2),p_{i}(t_2)\}=\cdots= \{x_{i}(t_j),p_{i}(t_j)\};\,t_1<t_2<\cdots<t_j,  \label{state self crossings}
\end{equation}%
as documented in figures 1(b), 2(b), 3(b), 4(b), 5(b), 6(b), 7(b), 8(b), 9(a), and 9(b). This is an indication that a classical state may experience instants of its past and/or future as it evolves in time..

Finally, the above methodical proposal may be extended into systems including the Duffing oscillators \cite{Bagchi Ghosh 2013,Agarwal}, Stochastic problems involving Langevin dynamical equation \cite{Kharkongor,Saikia,da Costa1 2020}, etc.


\begin{thebibliography}{99}
\bibitem{Symon 1972} K.R. Symon  (1972), \textit{Mechanics}, 3rd Edition
(Reading, MA: Addison-Wesley).

\bibitem{Chandrasekar 2006} V.K. Chandrasekar, M. Santhilvelan, A. Kundu, M.
Lakshmanan, J. Phys. \textbf{A}: Math. Gen. \textbf{39,} (2006) 9743.

\bibitem{Chandrasekar 2012} V.K. Chandrasekar, J.H. Sheeba, R.G. Pradeep,
R.S. Divyasree, M. Lakshmanan, Phys. Lett. \textbf{A 376} (2012) 2188.

\bibitem{Chandrasekar-PRE 2005} V.K. Chandrasekar, M. Santhilvelan, M.
Lakshmanan, Phys. Rev. E \textbf{72} (2005) 066203.

\bibitem{Carinena Ranada Sant 2004} J. F. Cari\~{n}ena, M. F. Ra\~{n}ada, M.
Santander, M. Senthilvelan, Nonlinearity \textbf{17} (2004) 1941.

\bibitem{Mustafa 2015} O. Mustafa, J. Phys. \textbf{A}; Math. Theor. \textbf{%
48} (2015) 225206.

\bibitem{Mustafa 2019} O. Mustafa, J Phys \textbf{A}: Math. Theor.\textbf{52
(}2019\textbf{) }148001.

\bibitem{Mustafa 2021} O. Mustafa, Eur. Phys. J. Plus \textbf{136} (2021)
249.

\bibitem{Mustafa1 2021} O. Mustafa, Phys. Scr. \textbf{96} (2021) 065205.

\bibitem{M-L 1974} P. M. Mathews, M. Lakshmanan, Quart. Appl. Math. \textbf{%
32 }(1974)\textbf{\ }215.

\bibitem{Tiwari 2013} A. K. Tiwari, S. N. Pandey, M. Santhilvelan, M.
Lakshmanan, J. Math. Phys. \textbf{54} (2013) 053506.

\bibitem{Lak-Chand 2013} M. Lakshmanan, V. K. Chandrasekar, Eur. Phys J. ST 
\textbf{222} (2013) 665.

\bibitem{Pradeep 2009} R. G. Pradeep, V. K. Chandrasekar, M. Santhilvelan,
M. Lakshmanan, J. Math. Phys. \textbf{50} (2009) 052901.

\bibitem{Chand-Lak 2007} V. K. Chandrasekar, M. Santhilvelan, M. Lakshmanan,
J. Math. Phys. \textbf{48} (2007) 032701.

\bibitem{Javier 2020} J. T. Marmolejo, O Isaksson, R C Trujillo, N C
Giesselmann, D Hanstorp, Am. J. Phys. \textbf{88 }(2020) 490.

\bibitem{Ortiz1} F. Olivar-Romero, O Rosas-Ortiz, J. Phys.: Conf. Ser. 
\textbf{839} (2017) 012010.

\bibitem{Ortiz2} F. Olivar-Romero, O Rosas-Ortiz, J. Phys.: Conf. Ser. 
\textbf{698} (2016) 012025.

\bibitem{Ortiz3} O. Rosas-Ortiz, N. F. Garcia, S. Cruz y Cruz, AIP Conf.
Proc. \textbf{1077} (2008) 31.

\bibitem{Kharkongor} D. Kharkongor, W. L. Reenbohn, M. C. Mahato, Phys. Rev.
E \textbf{94} (2016) 022148.

\bibitem{Saikia} S. Saikia, A. M. Jayannavar, M. C. Mahato, Phys. Rev. E 
\textbf{83} (2011) 061121.

\bibitem{da Costa1 2020} B. G. da Costa, I. S. Gomez, E. P. Borges, Phys.
Rev. E \textbf{102} (2020) 062105.

\bibitem{Khlevniuk 2018} A. Khlevniuk, V. Tymchyshyn, J. Math. Phys. \textbf{%
59} (2018) 082901.

\bibitem{Mustafa Algadhi 2019} O. Mustafa, Z. Algadhi, Eur. Phys. J. Plus 
\textbf{134} (2019) 228.

\bibitem{Quesne 2015} C. Quesne, J. Math. Phys. \textbf{56} (2015) 012903.

\bibitem{Mustafa Phys.Scr. 2020} O. Mustafa, Phys. Scr. \textbf{95} (2020)
065214.

\bibitem{Ranada 2016} M. Ranada, J. Math. Phys. \textbf{57}, 052703 (2016).

\bibitem{Carinena Herranz 2017} J. F. Cari\~{n}ena, F. J. Herranz, M. F. Ra%
\~{n}ada, J. Math. Phys. \textbf{58} (2017) 022701.

\bibitem{Bagchi Ghosh 2013} B. Bagchi, S. Das, S. Ghosh and S. Poria, J.
Phys. \textbf{A}: Math. Theor. \textbf{46,} (2013) 032001.

\bibitem{Mustafa 2020} O. Mustafa, Phys. Lett. \textbf{A 384} (2020) 126265.

\bibitem{dos Santos 2021} M. A. F. dos Santos, I. S. Gomez, B. G. da Costa,
O. Mustafa, Eur. Phys. J. Plus \textbf{136 }(2021) 96.

\bibitem{Mustafa arXiv} O. Mustafa, Eur. Phys. J. Plus \textbf{136 }(2021)
249.

\bibitem{Nabulsi1 2020} R. A. El-Nabulsi, Few-Body syst. \textbf{61} (2020)
37.

\bibitem{Nabulsi2 2020} R. A. El-Nabulsi, J. Phys. Chem.Solids \textbf{140}
(2020) 109384.

\bibitem{Nabulsi3 2021} R. A. El-Nabulsi, Physica E \textbf{134} (2021)
114827.

\bibitem{Nabulsi4 2021} R. A. El-Nabulsi, Optical and Quantum Electronics 
\textbf{53} (2021) 503.

\bibitem{Nabulsi5 2021} R. A. El-Nabulsi, Nanosystems. Nanostruct. \textbf{%
127} (2021) 114525.

\bibitem{Quesne 2004} C. Quesne, V. M. Tkachuk, J. Phys. \textbf{A 37}
(2004) 4267.

\bibitem{Quesne 2019} C. Quesne, Eur. Phys. J. Plus \textbf{134} (2019) 391.

\bibitem{Mustafa arXiv1} O. Mustafa, arXiv:2107.13758 " Modified Emden and the corresponding standard PDM dynamical equations: exact solvability and classical-states crossings".

\bibitem{Agarwal} V. Agarwal, X. Zheng, B. Balachandran, Phys. Lett. {A382} (2018) 3355.

\end{thebibliography}
\end{document}